\documentclass[twocolumn]{aastex631}
\usepackage{graphicx} 
\usepackage{amsmath}
\usepackage{booktabs}
\usepackage{mhchem}
\usepackage{gensymb}
\usepackage[T1]{fontenc}
\usepackage{floatrow} 
\usepackage{tablefootnote}
\usepackage{threeparttable}
\let\tablenum\relax
\usepackage{siunitx}
\usepackage[T1]{fontenc}

\shorttitle{Five New Small Planets}
\shortauthors{Gomez Barrientos et al.}

\begin{document}

\title{From Earths to Super-Earths: Five New Small Planets Transiting M Dwarf Stars}

\correspondingauthor{Jonathan Gomez Barrientos}
\email{jgomezba@caltech.edu}

\author[0000-0002-0672-9658]{Jonathan Gomez Barrientos}
\affiliation{Division of Geological and Planetary Sciences, California Institute of Technology, Pasadena, CA 91125, USA}

\author[0000-0002-5375-4725]{Heather A. Knutson}
\affiliation{Division of Geological and Planetary Sciences, California Institute of Technology, Pasadena, CA 91125, USA}

\author[0000-0001-9518-9691]{Morgan Saidel}
\altaffiliation{NSF Graduate Research Fellow}
\affiliation{Division of Geological and Planetary Sciences, California Institute of Technology, Pasadena, CA 91125, USA}

\author[0000-0002-0371-1647]{Michael Greklek-McKeon}
\affiliation{Earth and Planets Laboratory, Carnegie Institution for Science, 5241 Broad Branch Road NW, Washington, DC 20015, USA}
\affiliation{Division of Geological and Planetary Sciences, California Institute of Technology, Pasadena, CA 91125, USA}

\author[0000-0002-1422-4430]{W. Garrett Levine}
\affiliation{Department of Earth, Planetary, and Space Sciences, University of California, Los Angeles, CA 90095, USA}
\affiliation{Department of Astronomy, Yale University, New Haven, CT, 06511, USA}

\author[0000-0003-2657-3889]{Nicholas Saunders}
\affiliation{Department of Astronomy, Yale University, New Haven, CT, 06511, USA}
\affiliation{Institute for Astronomy, University of Hawai'i at M\=anoa, 2680 Woodlawn Drive, Honolulu, HI 96822, USA}

\author[0000-0002-0531-1073]{Howard Isaacson}
\affiliation{501 Campbell Hall, University of California at Berkeley, Berkeley, CA 94720, USA}

\author[0000-0003-2215-8485]{Renyu Hu}
\affiliation{Jet Propulsion Laboratory, California Institute of Technology, 4800 Oak Grove Drive, Pasadena, CA 91109, USA}
\affiliation{Division of Geological and Planetary Sciences, California Institute of Technology, Pasadena, CA 91125, USA}

\author[0000-0001-6588-9574]{Karen A.\ Collins}
\affiliation{Center for Astrophysics ${\rm \mid}$ Harvard {\rm \&} Smithsonian, 60 Garden Street, Cambridge, MA 02138, USA}

\author[0000-0002-5741-3047]{David R. Ciardi}
\affiliation{NASA Exoplanet Science Institute, IPAC, California Institute of Technology, Pasadena, CA 91125, USA}

\author[0009-0002-0733-572X]{Polina A. Budnikova}
\affiliation{Sternberg Astronomical Institute, M.V. Lomonosov Moscow State University, 13, Universitetskij pr., 119234, Moscow, Russia}

\author[0009-0003-4203-9667]{Dmitry V. Cheryasov}
\affiliation{Sternberg Astronomical Institute, M.V. Lomonosov Moscow State University, 13, Universitetskij pr., 119234, Moscow, Russia}

\author[0000-0001-7961-3907]{Samuel W. Yee}
\affiliation{Center for Astrophysics ${\rm \mid}$ Harvard {\rm \&} Smithsonian, 60 Garden Street, Cambridge, MA 02138, USA}
\affiliation{Department of Astrophysical Sciences, Princeton University, 4 Ivy Lane, Princeton, NJ 08544, USA}

\author[0000-0002-7883-5425]{Diogo Souto}
\affiliation{Departamento de F\'isica, Universidade Federal de Sergipe, Av. Marcelo Deda Chagas, S/N, 49107-230, S\~ao Crist\'ov\~ao, SE, Brazil}

\author[0000-0003-0012-9093]{Aida Behmard}
\affiliation{Center for Computational Astrophysics, Flatiron Institute, 162 Fifth Ave, New York, NY 10010, USA}

\author[0000-0002-4909-5763]{Akihiko Fukui}
\affiliation{Komaba Institute for Science, The University of Tokyo, 3-8-1 Komaba, Meguro, Tokyo 153-8902, Japan}
\affiliation{Instituto de Astrofisica de Canarias (IAC), 38205 La Laguna, Tenerife, Spain}

\author[0000-0002-1836-3120]{Avi Shporer}
\affiliation{Department of Physics and Kavli Institute for Astrophysics and Space Research, Massachusetts Institute of Technology, Cambridge, MA 02139, USA}

\author[0000-0003-0335-6435]{Akanksha Khandelwal}
\affiliation{Universidad Nacional Autónoma de México, Instituto de Astronomía, AP 70-264, Ciudad de M\'exico, 04510, México}

\author[0000-0001-8879-7138]{Bob Massey}
\affil{Villa '39 Observatory, Landers, CA 92285, USA}

\author[0000-0002-9355-5165]{Brice-Oliver Demory}
\affiliation{Center for Space and Habitability, University of Bern, Bern, Switzerland}
\affiliation{ARTORG Center for Biomedical Engineering Research, University of Bern, Bern, Switzerland}
\affiliation{Space Research and Planetary Sciences, Physics Institute, University of Bern, Bern, Switzerland}

\author[0000-0002-2361-5812]{Catherine A. Clark}
\affiliation{NASA Exoplanet Science Institute, IPAC, California Institute of Technology, Pasadena, CA 91125, USA}

\author[0000-0003-2163-1437]{Chris Stockdale}
\affiliation{Hazelwood Observatory, Australia}

\author[0000-0002-0388-8004]{Emily A. Gilbert}
\affiliation{NASA Exoplanet Science Institute, IPAC, California Institute of Technology, Pasadena, CA 91125, USA}

\author[0000-0003-0987-1593]{Enric Palle}
\affiliation{Instituto de Astrofisica de Canarias (IAC), 38205 La Laguna, Tenerife, Spain}
\affiliation{Departamento de Astrof\'\i sica, Universidad de La Laguna (ULL), 38206, La Laguna, Tenerife, Spain}

\author[0000-0003-2127-8952]{Francis P. Wilkin}
\affiliation{Department of Physics and Astronomy, Union College, 807 Union St., Schenectady, NY 12308, USA}

\author[0000-0001-9087-1245]{Felipe Murgas}
\affiliation{Instituto de Astrofisica de Canarias (IAC), 38205 La Laguna, Tenerife, Spain}
\affiliation{Departamento de Astrof\'\i sica, Universidad de La Laguna (ULL), 38206, La Laguna, Tenerife, Spain}

\author{Francis Zong Lang}
\affiliation{Center for Space and Habitability, University of Bern, Gesellschaftsstrasse 6, 3012, Bern, Switzerland}

\author{Ilse Plauchu-Frayn}
\affiliation{Universidad Nacional Autónoma de México, Instituto de Astronomía,  AP 106, Ensenada 22860, BC, México}

\author[0000-0002-8035-4778]{Jessie L. Christiansen}
\affiliation{NASA Exoplanet Science Institute, IPAC, California Institute of Technology, Pasadena, CA 91125, USA}

\author[0000-0002-4715-9460]{Jon M. Jenkins}
\affiliation{NASA Ames Research Center, Moffett Field, CA 94035, USA}

\author[0000-0002-6778-7552]{Joseph D. Twicken}
\affiliation{SETI Institute, Mountain View, CA 94043 USA/NASA Ames Research Center, Moffett Field, CA 94035 USA}
\affiliation{NASA Ames Research Center, Moffett Field, CA 94035, USA}

\author[0000-0003-1728-0304]{Keith Horne}
\affiliation{SUPA Physics and Astronomy, University of St. Andrews, Fife, KY16 9SS Scotland, UK}

\author[0000-0003-1462-7739]{Michaël Gillon}
\affiliation{Astrobiology Research Unit, Université de Liège, Allée du 6 Août 19C, B-4000 Liège, Belgium}

\author[0000-0001-9699-1459]{Monika Lendl}
\affiliation{Observatoire astronomique de l'Université de Genève, Chemin Pegasi 51, 1290 Versoix, Switzerland}

\author[0000-0003-2527-1598]{Michael B. Lund} 
\affiliation{NASA Exoplanet Science Institute, IPAC, California Institute of Technology, Pasadena, CA 91125, USA}

\author[0000-0001-8511-2981]{Norio Narita}
\affiliation{Komaba Institute for Science, The University of Tokyo, 3-8-1 Komaba, Meguro, Tokyo 153-8902, Japan}
\affiliation{Astrobiology Center, 2-21-1 Osawa, Mitaka, Tokyo 181-8588, Japan}
\affiliation{Instituto de Astrofisica de Canarias (IAC), 38205 La Laguna, Tenerife, Spain}

\author[0000-0002-4829-7101]{Pam Rowden}
\affiliation{Royal Astronomical Society, Burlington House, Piccadilly, London W1J 0BQ, UK}

 \author[0000-0003-3904-6754]{Ramotholo Sefako} 
\affiliation{South African Astronomical Observatory, P.O. Box 9, Observatory, Cape Town 7935, South Africa}

\author[0000-0001-8227-1020]{Richard P. Schwarz}
\affiliation{Center for Astrophysics ${\rm \mid}$ Harvard {\rm \&} Smithsonian, 60 Garden Street, Cambridge, MA 02138, USA}

\author[0000-0002-8965-3969]{Steven Giacalone}
\affiliation{Department of Astronomy, California Institute of Technology, Pasadena, CA 91125, USA}

\author[0009-0005-5046-203X]{Urs Schroffenegger}
\affiliation{Center for Space and Habitability, University of Bern, Gesellschaftsstrasse 6, 3012, Bern, Switzerland}

\author[0000-0002-7486-6726]{Yilen Gómez Maqueo Chew}
\affiliation{Universidad Nacional Autónoma de México, Instituto de Astronomía, AP 70-264, Ciudad de M\'exico, 04510, México}

\begin{abstract}

Earth-sized planets transiting M dwarf stars present one of the best opportunities with current facilities for studying the atmospheric and bulk compositions of terrestrial worlds. Here, we statistically validate five new transiting Earth and super-Earth sized planets orbiting M dwarf stars using a combination of light curves from the Transiting Exoplanet Survey Satellite, multi-color observations from Palomar and Las Cumbres Observatory, high-resolution imaging, and stellar spectroscopy. The sample includes: TOI-5716~b, an Earth-sized planet ($\mathrm{R_p}$ = 0.96 $\pm$ 0.05 $\mathrm{R_{\earth}}$) with a 6.766-day orbit around a metal-poor thin-disk star ([Fe/H] = $-0.54 \pm 0.10$); TOI-5728~b, a super-Earth ($\mathrm{R_p}$ = 1.31 $\pm$ 0.05 $\mathrm{R_{\earth}}$) on an 11.497-day orbit; and TOI-5736~b, a larger planet ($\mathrm{R_p}$ = 1.56 $\pm$ 0.07 $\mathrm{R_{\earth}}$) with an ultra-short period of just 0.649 days. We also statistically validate a multi-planet system, TOI-5489, hosting two similarly-sized super-Earths: TOI-5489~b ($\mathrm{R_p}$ = 1.40 $\pm$ 0.05 $\mathrm{R_{\earth}}$) and TOI-5489~c ($\mathrm{R_p}$ = 1.28 $\pm$ 0.07 $\mathrm{R_{\earth}}$) with orbital periods of 3.152 and 4.921 days, respectively. Due to their longer orbital periods, TOI-5716~b and TOI-5728~b both have equilibrium temperatures $\leq$ 400~K, making them useful test cases for studies of atmospheric mass loss. If TOI-5728~b is confirmed to have an Earth-like bulk composition, it would join the very small sample of rocky planets orbiting mid- to late-M dwarfs that lie below the cosmic shoreline and therefore may have retained high mean molecular weight atmospheres.

\end{abstract}

\section{Introduction}
\label{sec:intro}

A primary goal of exoplanetary science is to detect and characterize Earth-sized exoplanets, particularly those orbiting within or near the habitable zones of their host stars. While detecting and characterizing these worlds around Sun-like stars presents significant challenges, some of these challenges can be mitigated by concentrating on transiting Earth-sized planets that orbit M-type stars -- the most abundant stellar type in the Milky Way \citep[][]{Henry1994, Kirkpatrick1999}. The smaller radii of M dwarfs result in deeper and more frequent planetary transits compared to similar planets orbiting Sun-like stars, making them more favorable targets for atmospheric characterization via transmission spectroscopy. The lower masses of M dwarfs also result in larger radial velocity semi-amplitudes, making planets around these stars more favorable targets for radial velocity mass measurements.  For these reasons, small planets around M dwarfs offer an unprecedented opportunity to characterize the bulk densities and atmospheric or surface properties of terrestrial worlds using current ground-based and space-based telescopes \citep[e.g.,][]{Nutzman2008, Wordsworth2021}.

However, despite these observational advantages, M dwarfs also present unique challenges for atmospheric characterization studies. Although it is easier to detect atmospheric signals from Earth-sized planets transiting M dwarfs, they may be less likely to retain their primary atmospheres than their counterparts around Sun-like stars \citep{Hori2020}. M dwarfs emit more frequent flares and have higher fractional X-ray and extreme ultraviolet (XUV) fluxes, which can more efficiently strip away the gaseous envelopes of planets with equivalent equilibrium temperatures \citep{Roettenbacher2017, Fleming2020}. There is abundant observational evidence indicating that most small planets on close-in orbits around M dwarfs lack thick atmospheres \citep[e.g.,][]{Greene2023, Zieba2023, Zhang2024, Xue2024, Wachiraphan2025, MeierValdes2025, Luque2025, Fortune2025, Kreidberg2025}, but current surveys have not yet been able to determine the minimum orbital separation outside which rocky planets consistently retain high mean molecular weight atmospheres. 

The concept of the ``cosmic shoreline'' was first introduced by \citet{Zahnle2017}, who observed that in the Solar System it is possible to draw a line in total insolation versus escape velocity that separates rocky bodies with and without atmospheres. While it is the XUV irradation that drives atmospheric mass loss, this study pointed out that total insolation can be converted to an equivalent XUV irradiation for FGK and early M stars using a simple power law scaling.  More recently, \citet{Pass2025} extended this framework to mid-to-late M stars, which exhibit unique activity histories compared to early M stars and FGK stars and therefore require a different approach to estimate their time-integrated XUV fluxes.

Our ability to observationally test the cosmic shoreline framework for both early and mid-to-late M dwarfs is currently limited by the small number of confirmed rocky planets with low equilibrium temperatures transiting nearby M dwarfs. To this end, we statistically validate the planetary nature of five small ($<1.8$ R$_\earth$) TESS planet candidates transiting early-to-mid M-dwarfs ($T_\mathrm{eff}$ of $3300-3900$~K): TOI-5489.01, TOI-5489.02, TOI-5716.01, TOI-5728.01, and TOI-5736.01. Four of these candidates have predicted equilibrium temperatures lower than 650~K, and two are $\leq$ 400~K, making them useful for constraining the location of the cosmic shoreline.  We begin by describing the observations of our targets in Section~\ref{sec:obs}, including the TESS photometry, multi-color ground-based photometry, high-resolution imaging, and stellar spectroscopy. Our light curve modeling procedure is described in Section~\ref{sec:modeling} and our statistical validation analysis in detailed in Section~\ref{sec:validation}. We present our results in Section~\ref{sec:results} and discuss their implications in Section \ref{sec:discussion}. Finally, we summarize our conclusions in Section~\ref{sec:conc}.

\section{Observations}
\label{sec:obs}

\begin{table*}[]
\centering
\fontsize{9pt}{9pt}\selectfont
\begin{tabular}{lllll}
\hline
Parameter & TOI-5489$^a$ & TOI-5716$^b$ & TOI-5728$^c$ & TOI-5736$^a$ \\ 
\hline
TIC ID & 468983280 & 298074836 & 219875976 & 310380289 \\
RA & 08:34:45.91 & 13:27:36.19 & 17:21:24.22 & 14:15:02.02 \\
Dec & +11:29:22.52 & +73:10:38.76 & +67:33:01.52 & +49:30:03.38 \\
$TESS$ (mag) & 11.69 & 13.10 & 12.49 & 11.98 \\
$V$ (mag) & 13.82 & 15.83 & 15.02 & 13.55 \\
$J$ (mag) & 10.28 & 11.57 & 11.00 & 10.83 \\
$\mathrm{T_{eff}}$ (K) & $3547 \pm 157$ & $3436 \pm 79$ & $3419 \pm 70$ & $3949 \pm 157$ \\
$\mathrm{[Fe/H]}$ & -- & $-0.54 \pm 0.10$ & $-0.12 \pm 0.09$ & -- \\
$\mathrm{R_*}$ ($\mathrm{R_\odot}$) & $0.42 \pm 0.01$ & $0.22 \pm 0.01$ & $0.37 \pm 0.01$ & $0.58 \pm 0.02$ \\
$\mathrm{M_*}$ ($\mathrm{M_\odot}$) &  $0.41 \pm 0.02$ & $0.19 \pm 0.01$ & $0.36 \pm 0.02$ & $0.57 \pm 0.02$ \\
Plx (mas) & $22.36 \pm 0.02$ & $25.41 \pm 0.02$ & $18.23 \pm 0.02$ & $10.82 \pm 0.01$ \\
RV (km/s) & $37.71 \pm 0.89^e$ & $25.50 \pm 0.10^d$ & $-6.81 \pm 0.10^d$ &  $-1.76 \pm 0.99^e$    \\
\hline
\end{tabular}
\caption{Summary of stellar properties}
\label{tab:stellar_props}
\vspace{0.1cm}
\footnotesize
\begin{flushleft}
\textbf{Notes.}
\tablenotemark{a} Stellar properties originate from the TESS Input Catalog (TIC; \citealt{Stassun2019}). \\
\tablenotemark{b} $T_\mathrm{eff}$, and [Fe/H] were derived from APOGEE spectra as described in Section \ref{sec:apogee}. $\mathrm{R_*}$ and $\mathrm{M_*}$ were derived using \texttt{isoclassify}. All other values are from the TIC. \\
\tablenotemark{c} $T_\mathrm{eff}$, and [Fe/H] were derived from HIRES spectra using the SpecMatch-Emp tool \citep{Yee2017} as described in Section \ref{sec:spectra}. $\mathrm{R_*}$ and $\mathrm{M_*}$ were derived using \texttt{isoclassify}. All other values are from the TIC. \\
\tablenotemark{d} From Keck/HIRES.
\tablenotemark{e} From GAIA DR3 \citep[e.g.,][]{Katz2023}.
\end{flushleft}
\end{table*}

\begin{table*}[h]
\centering
\fontsize{8pt}{8pt}\selectfont
\begin{tabular}{llllllllll}
\hline
TOI & Observatory & $r_\mathrm{phot}$\tablenotemark{\scriptsize a} & Date  & Start Time & End Time & Filter(s) & $t_{\exp}$ & Airmass & $\sigma/\sigma_\mathrm{phot}$\tablenotemark{\scriptsize c} \\
 &  & (\arcsec) & (UTC) &  &  &  & (s) & Start/Middle/End &  \\
\hline
5489.01 & LCO-McD & 5.06 & 2022-12-11 & 05:47:19 & 08:53:25 & $i'$ & 37 & 2.27/1.39/1.12 & 1.2 \\
5489.01 & LCO-McD & 7.39 & 2023-02-09 & 03:03:03 & 06:23:41 & $i'$ & 37 & 1.51/1.14/1.06 & 1.1 \\
5489.01 & LCO-CTIO & 5.84 & 2023-02-09 & 03:02:54 & 06:23:56 & $i'$ & 37 & 1.39/1.36/1.68 & 1.5   \\
5489.01 & LCO-SSO & 5.45 & 2023-02-15 & 10:21:09 & 13:42:25 & $i'$ & 37 & 1.84/1.41/1.39 & 1.4  \\
5489.01 & LCO-SAAO & 5.45 & 2023-02-21 & 18:25:55 & 21:02:32 & $i'$ & 37 & 1.90/1.50/1.39 & 2.0 \\
5489.01 & LCO-Teid & 4.67 & 2024-01-02 & 22:56:18 & 02:19:42 & $i'$ & 37 & 1.87/1.22/1.06 & 1.4  \\
5489.02 & LCO-SAAO & 3.89 & 2025-01-01 & 20:50:15 & 00:33:34 & $i'$ & 37 & 2.70/1.54/1.39 & 2.6  \\
5489.02 & LCO-McD & 3.89 & 2025-02-04 & 07:34:11 & 11:04:42 & $i'$ & 37 & 1.09/1.35/2.38 & 1.5  \\
5489.02 & LCO-CTIO & 6.61 & 2025-04-28 & 23:35:20 & 02:41:33 & $i'$ & 37 & 1.37/1.66/2.82 & 1.4  \\
5489.02 & Palomar-Hale & 3.75 & 2024-12-07 & 07:14:35 & 09:56:44 & $J$ & 28 & 2.02/1.39/1.15 & 4.9  \\
5716.01 & Palomar-Hale & 4.00 & 2025-01-24 & 08:57:07 & 14:20:58 & $J$ & 36 & 1.55/1.32/1.32 & 3.3 \\
5716.01 & SAINTEX & 4.20 & 2025-01-24 & 09:24:29 & 14:00:18 & $I+z$ & 46 & 1.54/1.38/1.36 & 1.0 \\
5716.01 & SAINTEX & 4.55 & 2025-04-22 & 07:01:17 & 12:28:54 & $I+z$ & 50 & 1.34/1.52/1.86 & 1.0  \\
5728.01 & Palomar-Hale & 4.00 & 2024-10-10 & 02:26:09 & 06:20:00 & $J$ & 40 & 1.32/1.59/2.13 & 3.5 \\
5736.01 & Palomar-Hale & 3.50 & 2024-07-28 & 03:50:32 & 07:07:32 & $J$ & 30 & 1.15/1.42/2.06 & 3.8  \\
5736.01 & LCO-FTN & 7.56 & 2024-02-19 & 10:34:07 & 16:01:09 & $g',r', i', z_s$ & 60/23/23/27 & 1.82/1.20/1.18 & 1.2/1.4/1.5/1.6\\
5736.01\tablenotemark{\scriptsize b} & LCO-FTN & 5.67 & 2024-03-27 & 10:42:15 & 15:30:03 & $g',r', i', z_s$ & 60/23/23/27 & 1.21/1.15/1.47 & 1.1/1.4/1.4/1.6 \\
5736.01 & LCO-FTN & 7.83 & 2025-01-02 & 12:53:12 & 15:47:25 & $g',r', i', z_s$ & 60/23/23/27 & 2.32/1.57/1.26 & 1.4/2.0/1.9/2.1 \\
5736.01 & LCO-FTN & 7.29 & 2025-05-14 & 05:51:54 & 09:01:41 & $g',r', i', z_s$ & 60/23/23/27 & 1.52/1.24/1.14 & 1.1/1.4/1.5/1.6\\
5736.01 & LCO-Teid & 5.06 & 2024-03-04 & 01:36:54 & 06:28:57 & $i'$ & 42 & 1.30/1.08/1.16 & 1.0 \\
5736.01 & LCO-McD & 4.28 & 2024-04-02 & 06:52:13 & 11:46:23 & $i'$ & 42 & 1.11/1.07/1.34 & 1.5 \\
\hline
\end{tabular}
\caption{Ground-based light curve observation details.}. 
\label{tab:lco_keplercam_obs}
\vspace{0.05cm}
\footnotesize
\begin{flushleft}
\textbf{Notes.}
\tablenotemark{a} Radius of optimal photometric aperture. \tablenotemark{b} The observations in $r', i',$ and $z_s$ exhibit correlated noise and were excluded from the global fit. 
\tablenotemark{c} The true uncertainty in the data divided by the photon noise.
\end{flushleft}
\end{table*}

\begin{table*}[h!]
\fontsize{9pt}{9pt}\selectfont
\centering
\begin{tabular}{lllll}
\hline
TOI & Telescope & Instrument & Filter & Image Type \\ \hline
5489 & & & & \\ \hline
& WIYN (3.5 m) & NESSI & 832 (40) nm & Speckle \\ \hline
& WIYN (3.5 m) & NESSI & 562 (44) nm & Speckle \\ \hline
& SAI-2.5m (2.5 m) & Speckle Polarimeter & I & Speckle \\ \hline
& \textbf{Palomar (5 m)} & PHARO & Kcont & AO \\ \hline
& SOAR (4.1 m) & HRCam & I & Speckle \\ \hline
5716 & & & & \\ \hline
& \textbf{SAI-2.5m} (2.5 m) & Speckle Polarimeter & I & Speckle \\ \hline
5728 & & & & \\ \hline
& WIYN (3.5 m) & NESSI & 562 (44) nm & Speckle \\ \hline
& \textbf{Keck2 (10 m)} & NIRC2 & Kcont & AO \\ \hline
& SAI-2.5m (2.5 m) & Speckle Polarimeter & I & Speckle \\ \hline
& WIYN (3.5 m) & NESSI & 832 (40) nm & Speckle \\ \hline
5736 & & & & \\ \hline
& Gemini (8 m) & 'Alopeke & 832 (40) nm & Speckle \\ \hline
& Gemini (8 m) & 'Alopeke & 562 (54) nm & Speckle \\ \hline
& WIYN (3.5 m) & NESSI & 832 (40) nm & Speckle \\ \hline
& WIYN (3.5 m) & NESSI & 562 (44) nm & Speckle \\ \hline
& \textbf{Keck2 (10 m)} & NIRC2 & Kcont & AO \\ \hline
& SAI-2.5m (2.5 m) & Speckle Polarimeter & I & Speckle \\ \hline
\end{tabular}
\caption{Details of high-resolution imaging. The observations adopted in our statistical validation analysis are marked in bold.}
\label{tab:imaging}
\end{table*}

\subsection{TESS Photometry}

All of our targets were initially identified by TESS through the detection of their periodic transit signals. In this work, we utilized all available 2-minute cadence Presearch Data Conditioning Simple Aperture Photometry (PDCSAP; \citet{Stumpe2012, Smith2012, Stumpe2014}) light curves from the TESS Science Processing Operations Center (SPOC) pipeline \citep{Jenkins2016}. We accessed the publicly available photometry for our five TESS candidates (TOI-5489.01, TOI-5489.02, TOI-5716.01, TOI-5728.01, and TOI-5736.01) through the Mikulski Archive for Space Telescope (MAST) using the Python package \texttt{lightkurve} \citep{2018ascl.soft12013L}. Prior to their release as TOI alerts, all candidates had been vetted by the TESS Science Office using automated search pipelines \citep{Guerrero2021}.

To ensure optimal photometric quality for subsequent transit analysis, we performed additional processing on the retrieved TESS observations. We first masked the transit signals and applied a Lomb-Scargle periodogram \citep{Lomb1976, Scargle1982} to identify any residual periodic trends that might impact the quality of individual transits. This analysis revealed systematic trends in all five targets' TESS light curves, with periods ranging from 4.8 to 9.6 days (TOI-5489: 8.4 d; TOI-5716: 9.6 d; TOI-5728: 4.8 d; TOI-5736: 6.5 d). To address these systematic effects, we modeled the residual periodicity using a Gaussian kernel convolution following the methodology described in \citet{Greklek-McKeon2023}, then divided out these trends to produce flattened TESS light curves. 

\subsection{Multi-Color Transit Photometry}
\subsubsection{Palomar Photometry}

We obtained $J$ band transit observations of four TESS candidates (TOI-5489.02, TOI-5716.01, TOI-5728.01, and TOI-5736.01) using the Wide-field InfraRed Camera (WIRC; \citet[][]{Wilson2003}) on the 200-inch Hale Telescope at Palomar Observatory. These targets were observed through a combination of dedicated telescope time and downtime allocated to other programs. All observations employed a beam-shaping diffuser that produces a top-hat point spread function (PSF) with a full width at half-maximum of 3\arcsec. This diffuser provided two key advantages: it stabilized the PSFs to reduce correlated noise from changing atmospheric conditions \citep{Stefansson2017, Vissapragada2020}, and it increased observing efficiency by enabling longer exposure times for bright targets. In Table \ref{tab:lco_keplercam_obs} we provide a summary of the observations.

Next, we extracted light curves for each target using the same general approach as in previous studies \citep{Vissapragada2020,Greklek-McKeon2023}. We first flat-fielded, dark-subtracted, and corrected the images for hot and dead pixels following the procedures detailed in \citet[][]{Vissapragada2020}. We then performed circular aperture photometry using the Python package \texttt{photutils} on all stars identified by the DAOStarFinder algorithm \citep{Stetson1987}. This algorithm detects stars that meet or exceed a manually specified signal-to-noise ratio threshold, which we set between 100 and 400 depending on target brightness and field star density.

To optimize the photometric precision, we tested aperture radii ranging from 5 to 30 pixels in steps of 1 pixel for each target. For each aperture size, we extracted light curves for both the target and up to 10 comparison stars, then detrended the target light curve by dividing it by the mean light curve of all comparison stars. We calculated the root mean square (rms) of the detrended data and selected the aperture size that minimized the target's light curve rms for all subsequent analyses.

\subsubsection{LCOGT Photometry}

We obtained multi-color transit observations of three TESS candidates (TOI-5489.01, TOI-5489.02, and TOI-5736.01) with nodes from the Las Cumbres Observatory Global Telescope network \citep[LCOGT;][]{Brown:2013}, including the 1\,m facilities at Teide Observatory, Tenerife (Teide) and McDonald Observatory, Fort Davis, Texas (McD), plus the 2\,m facility at Faulkes Telescope North (FTN), Haleakala Observatory, Maui, Hawai'i. Each 1\,m telescope houses a $4096\times4096$ SINISTRO camera with $0\farcs389$ per pixel resolution, providing a $26\arcmin\times26\arcmin$ field of view. The 2\,m telescope employs the MuSCAT3 multi-band imager \citep{Narita:2020}. We processed all LCOGT data using the standard {\tt BANZAI} calibration pipeline \citep{McCully:2018} and performed differential photometry using {\tt AstroImageJ} \citep{Collins:2017}. Details of the observations are provided in Table~\ref{tab:lco_keplercam_obs}.

\subsubsection{SAINT-EX Photometry}

We obtained \( I+z \) band transit observations of TOI-5716.01 using the 1.0-m SAINT-EX (Search And characterIsatioN of Transiting EXoplanets) telescope \citep{Demory2020, yilen2023}, located at the Sierra de San Pedro Mártir Observatory in Baja California, México. SAINT-EX is a Ritchey-Chrétien f/8 telescope equipped with a thermoelectrically cooled \( 2048 \times 2048 \) Andor iKon-L CCD camera, providing a pixel scale of \( 0\farcs35 \) and a field of view of \( 12\arcmin \times 12\arcmin \). Observations were conducted on 2025 January 24 and 2025 April 22 (UT). We reduced and analyzed the photometric data using {\tt AstroImageJ} \citep{Collins:2017}. A summary of the observations is provided in Table~\ref{tab:lco_keplercam_obs}.

\subsection{High Angular Resolution Imaging}

\begin{figure*}
    \centering
    \includegraphics[width=0.75\linewidth]{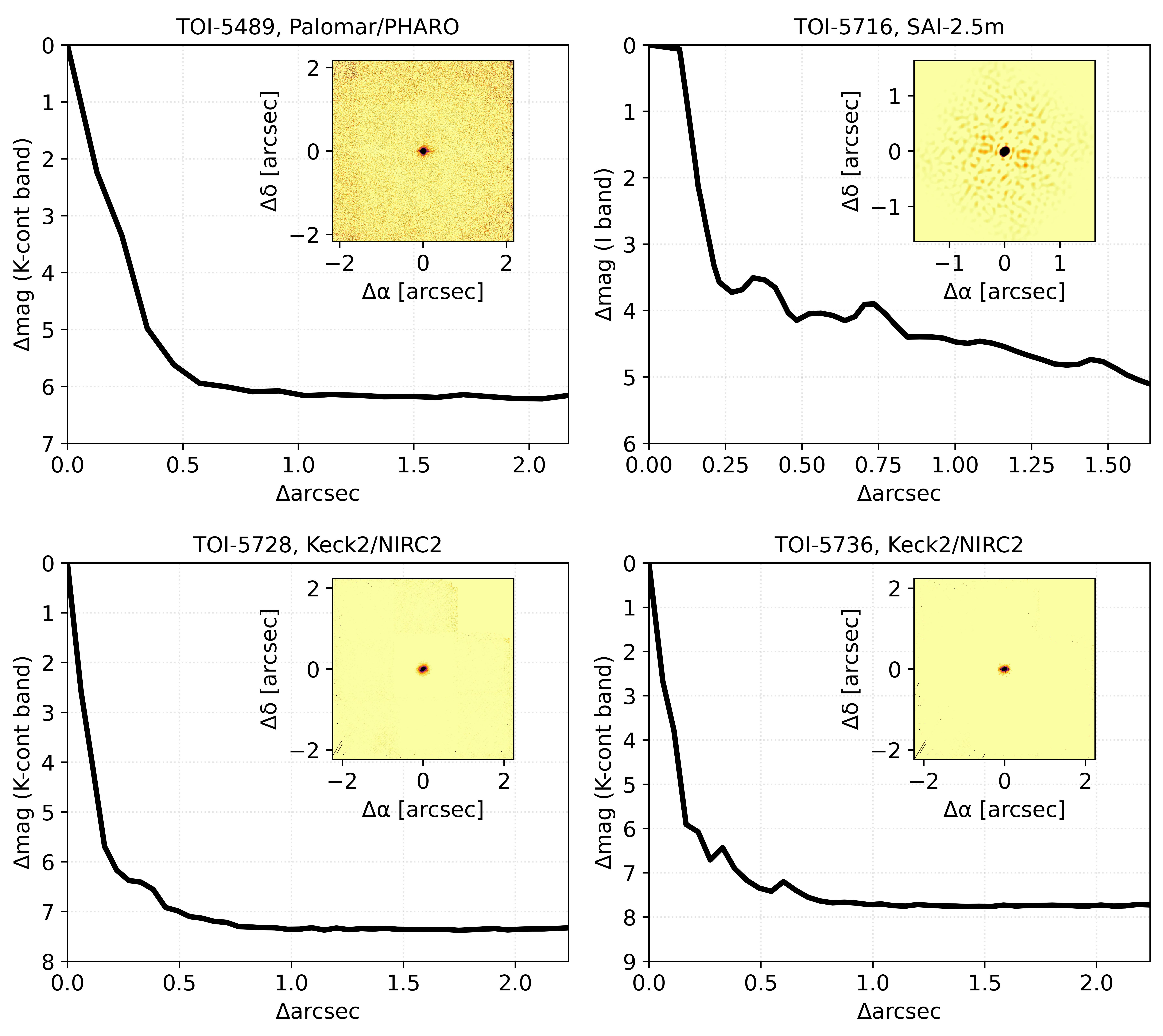}
    \caption{High-resolution images and corresponding contrast curves (black line) of our targets. There are no secondary sources detected around any of the targets.}
    \label{fig:ccs}
\end{figure*}

We downloaded all available high-resolution images and corresponding contrast curves for each target from the ExoFOP website\footnote{\url{https://exofop.ipac.caltech.edu/tess/}} to check for unresolved stellar companions that may potentially dilute or host the observed transit signals. These observations revealed no nearby stellar companions for any of our five targets (see Figure \ref{fig:ccs}). For targets with multiple contrast curves, we selected the observation with the best contrast sensitivity (see Table \ref{tab:imaging}) for our subsequent statistical validation analysis with \texttt{TRICERATOPS+} \citep{GomezBarrientos2025}.

\subsubsection{SAI-2.5m}

TOI-5716 was observed on 2023-10-16 UT, with the speckle polarimeter on the 2.5-m telescope at the Caucasian Observatory of Sternberg Astronomical Institute (SAI) of Lomonosov Moscow State University. A low--noise CMOS detector Hamamatsu ORCA--quest was used as a detector. The atmospheric dispersion compensator was active, which allowed the use of the $I_\mathrm{c}$ band. The respective angular resolution is $0.083\arcsec$. 2500 frames with exposure 60~ms were accumulated. No companion was detected.  We calculated standard deviation in concentric ring zones within the autocorrelation function (see inset in the upper right panel of Figure~\ref{fig:ccs}). This standard deviation multiplied by a factor of 5 was converted to the expected magnitude difference between the host star and a possible stellar companion \citep{Strakhov2023}. The detection limits at distances $0.25\arcsec$ and $1.0\arcsec$ from the star are $\Delta I_\mathrm{c}=3.8^m$ and $4.5^m$.

\subsubsection{Palomar and Keck2/NIRC2}

TOI~5489 was observed on 2023-Nov-29 UT with the PHARO instrument \citep{hayward2001} on the Palomar Hale (5m) behind the P3K natural guide star AO system \citep{dekany2013}. The pixel scale for PHARO is $0.025\arcsec$. The Palomar data were collected in a standard 5-point quincunx dither pattern. Observations of TOI-5728 and TOI-5736 were made on 2023-06-10 UT with the NIRC2 instrument on Keck-II (10m) behind the natural guide star AO system \citep{wizinowich2000} in the standard 3-point dither pattern that is used with NIRC2 to avoid the left lower quadrant of the detector.  All observations were made in the K$_{cont}$ filter. Flat fields were taken on-sky, dark-subtracted, and median averaged, and sky frames were generated from the median average of the dithered science frames. Each science image was then sky-subtracted and flat-fielded.  The reduced science frames were combined into a single mosaiced image with final resolutions of $\sim 0.11$\arcsec, $\sim 0.056$\arcsec, and $\sim 0.056$\arcsec, respectively.  The sensitivity of the final combined AO images were determined by injecting simulated sources azimuthally around the primary target every $20^\circ $ at separations of integer multiples of the central source's FWHM \citep{furlan2017}. The brightness of each injected source was scaled until standard aperture photometry detected it with $5\sigma$ significance.  The final $5\sigma$ limit at each separation was determined from the average of all of the determined limits at that separation and the uncertainty on the limit was set by the rms dispersion of the azimuthal slices at a given radial distance. No stellar companions were detected.

\subsection{High Resolution Spectroscopy}
\label{sec:spectra}
\subsubsection{HIRES Spectra}

We obtained observations of TOI-5716 and TOI-5728 with the High Resolution Echelle Spectrometer (HIRES; \citet{Vogt1994}) on the Keck I Telescope. For each target, we collected one out-of-transit, iodine-free spectrum on 2025-06-08 UT adopting an exposure time of 1800 seconds for TOI-5716 and 1200 seconds for TOI-5728. We provide the measured radial velocities in Table \ref{tab:stellar_props}. To empirically derive spectroscopic stellar parameters, we used the SpecMatch-Emp routine \citep{Yee2017}, which cross-matches a given HIRES spectrum with a library of high signal-to-noise observed spectra of stars with empirically determined stellar parameters. The technique is especially effective for late type stars for which synthetic spectra are limited due to incomplete line lists. 

For TOI-5728 we find [Fe/H] = -0.12 $\pm$ 0.09 and $\mathrm{T_{eff}}$ = 3419 $\pm$ {70}~K, which is in agreement with the TESS Input Catalog (TIC; \citet{Stassun2019}) value of $\mathrm{T_{eff}}$ = 3444 $\pm$ 157~K. For TOI-5716 we find [Fe/H] = -0.45 $\pm$ 0.09 and $\mathrm{T_{eff}}$ = 3269 $\pm$ {70}~K, which is also in agreement with the TIC value of $\mathrm{T_{eff}}$ = 3331 $\pm$ 157~K. We note that [Fe/H] = -0.45 lies at the lower boundary of the spectral library used by SpecMatch-Emp, and conclude that the true metallicity of TOI-5716 may be lower than this value. 

For TOI-5728, we derive $\mathrm{R_*}$ and $\mathrm{M_*}$ with the Python package \texttt{isoclassify} \citep{Huber2017, Berger2020, Berger2023}. The package incorporates spectroscopic information (e.g., $\mathrm{T_{eff}}$ and [Fe/H]), photometry (e.g., $K$ magnitude), and \textit{Gaia} parallax to derive stellar parameters. In our calculation, we selected the MESA Isochrones and Stellar Tracks \citep{Choi2016} and adopted the default settings. The derived values are given in Table \ref{tab:stellar_props}. We note that these values are consistent with the TIC values of $\mathrm{R_*}$ = 0.376 $\pm$ 0.011 and $\mathrm{M_*}$ = 0.364 $\pm$ 0.020, which are calculated using the $K_S$-band magnitude relations from \citet{Mann2015} and \citet{Mann2019}.

\subsubsection{APOGEE Spectra}
\label{sec:apogee}

We analyzed the H-band (1.51 -- 1.60 \micron) APOGEE high-resolution (\citealp{Wilson2018}; R$\sim$ 22,500) spectra of TOI-5716 to determine precise stellar parameters and individual abundances. APOGEE is a cryogenic multi-fiber spectrograph located in both hemispheres (\citealp{sloandigitalskysurveyv2025}) with ongoing observations for the Milky Way Mapper project within the SDSS-V survey. We used the normalized public APOGEE DR19 spectra (\citealt{Nidever_2015}; \citealt{sdsscollaboration2025}) in this study.

Previous studies have utilized APOGEE data to measure the metallicity of TOI-5716, though with varying results and limitations. TOI-5716 is included in the APOGEE DR16 catalog \citep{Ahumada2020}, with a reported metallicity of [m/H] = -1. However, \citet{Birky2020} and \citet{Souto2021} both found that this catalog is biased toward reporting higher temperatures and lower metallicities for M dwarfs. \citet{Birky2020} instead derive their own catalog using \emph{The Cannon}, a machine learning algorithm for spectral classification \citep{Ness2015, Casey2016}, and report [m/H] = $-0.77 \pm 0.08$ for TOI-5716. However, as shown in Figure 2 of \citet{Birky2020}, this metallicity lies outside the range of their training data, which only extends to [m/H] = -0.5, and therefore may also suffer from edge effects.

\begin{figure}
    \centering
    \includegraphics[width=0.99\columnwidth]{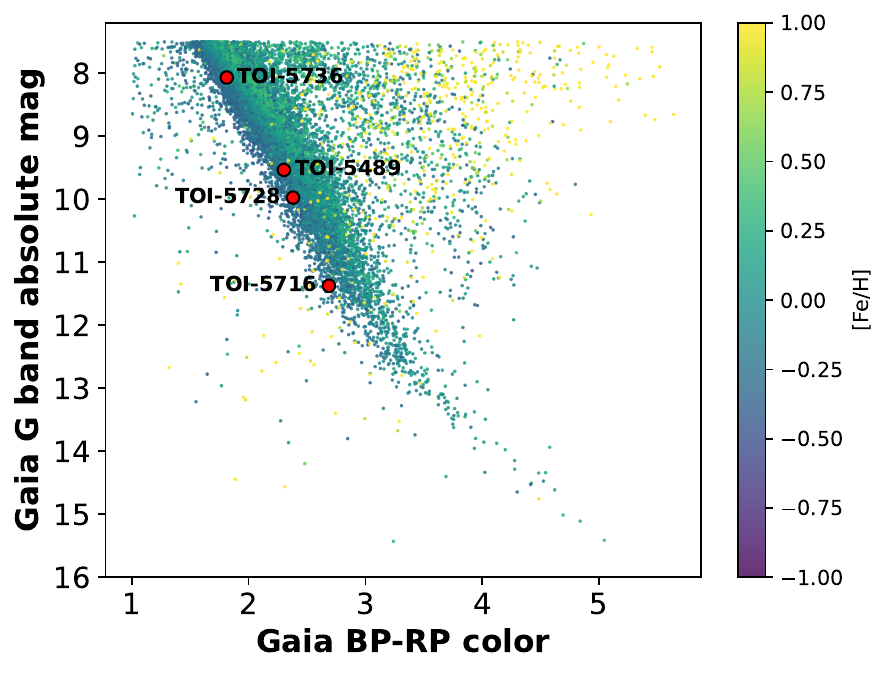}
    \caption{Gaia color-magnitude diagram of our targets, along with the catalog of M stars from \citealt{Birky2020}. The data points are color-coded by their metallicty from \citealt{Birky2020}. The location of TOI-5716 is consistent with the most metal-poor subset of this population, in good agreement with our spectroscopic analysis.}
    \label{fig:cmd}
\end{figure}

Given the limitations surrounding SpecMatch-Emp and previous work in deriving reliable spectroscopic parameters for TOI-5716, we performed a detailed analysis based on the high-resolution near-infrared spectra from APOGEE. M dwarf atmospheric parameters and elemental abundances are more effectively constrained in the near-infrared (NIR), where these stars are significantly brighter and exhibit fewer and less complex molecular features compared to the optical regime \citep{allard2000}. We adopted the methodology described in \citet{Souto2020} to derive the stellar atmospheric parameters ($T_{\rm eff}$, $\log g$) and individual abundances of C, O, Mg, Al, K, Ca, and Ti (see \citealt{Melo2024}). Using this framework, we derived values of $T_{\rm eff}$ = 3436 $\pm$ 79 K and log $g$ = 4.9 $\pm$ 0.13 dex for TOI-5716. This effective temperature is slightly higher than the one derived using HIRES ($3269\pm70$~K), but still consistent at the 1.6$\sigma$ level.  We obtain [Fe/H] = -0.54 $\pm$ 0.10 dex from our fit to the APOGEE spectra, in good agreement with the [Fe/H] = -0.45 $\pm$ 0.09 value estimated using HIRES spectra. We further verify this result by plotting TOI-5716, along with our other targets, in a color-magnitude diagram. As shown in Figure \ref{fig:cmd}, TOI-5716 lies below the solar-metallicity locus. Lastly, using the APOGEE spectra, we also constrain the elemental abundances of the following species: [C/Fe] = 0.07 $\pm$ 0.04, [O/Fe] = 0.13 $\pm$ 0.04, [Mg/Fe] = -0.04 $\pm$ 0.13, [Al/Fe] = -0.24 $\pm$ 0.11, [K/Fe] = 0.07 $\pm$ 0.08, [Ca/Fe] = 0.13 $\pm$ 0.08, [Ti/Fe] = -0.13 $\pm$ 0.10. Having obtained reliable spectroscopic parameters for TOI-5716, we followed the same approach as with TOI-5728 to derive $\mathrm{R_*}$ and $\mathrm{M_*}$ and report our results in Table \ref{tab:stellar_props}. Similar to TOI-5728, our derived values are consistent with the TIC values of $\mathrm{R_*}$ = 0.226 $\pm$ 0.007 and $\mathrm{M_*}$ = 0.195 $\pm$ 0.020.

\section{Light Curve Modeling}
\label{sec:modeling}

We used our photometric transit data to characterize the properties of each TESS candidate and investigate potential wavelength-dependent transit depth variations. To do so, we first fit the individual TESS transits, then phase-folded the TESS data on the best-fit period, and finally performed a joint fit of the phase-folded TESS data with the ground-based light curves.

We modeled the light curves of our targets using the \texttt{exoplanet} package, which couples the forward light curve model \texttt{starry} \citep{Luger2019} with \texttt{PyMC3's} No U-turn Sampler \citep[NUTS;][]{Hoffman2011} to explore the parameter space. In all fits, we simultaneously modeled both the transit signal and systematic effects in the ground-based light curves. For the transit model, we assumed circular orbits (i.e., zero orbital eccentricity) for all planets. We tested the validity of this assumption for TOI-5716~b and TOI-5728~b, which have the longest orbital periods in our sample, by repeating our fits with $e$cos$\omega$ and $e$sin$\omega$ as free parameters and placing a prior on the eccentricity $e$ following \citet{Kipping2013}. We found that the retrieved posteriors for $e$ are consistent with circular orbits for both planets, and that the data provide relatively weak constraints on this parameter. We therefore fixed $e=0$ in our final version of the fits for all five planets. We also adopted a quadratic limb darkening law with fixed coefficients calculated using \texttt{ExoTIC-LD} \citep{Grant2024} for each bandpass, based on the stellar properties reported in Table \ref{tab:stellar_props}. We also explored fits where we allowed the limb darkening coefficients to vary as free parameters with either broad uniform priors or normal priors centered on the predicted values with widths calculated by propagating the uncertainties in the stellar parameters. We found that this yielded equivalent $\mathrm{R_p/R_*}$ posteriors, indicating that the precision of our light curves is not high enough for us to be sensitive to these effects.

The transit model included four global parameters: the semi-major axis in stellar radius units ($a/R_*$), the impact parameter ($b$), the transit epoch ($t_0$), and the orbital period ($P$). For joint fits of phase-folded TESS data and ground-based light curves, we treated $P$ as a free parameter with a Gaussian prior centered on the TESS-only fit value and bounded by its uncertainties. We also fit for separate planet-to-star radius ratios ($R_p/R_*$) in each bandpass to search for potential wavelength-dependent depth variations. For the TESS data, we included a parameter ($k$) that scaled the reported TESS error bars to match the observed scatter in the residuals (i.e., $\sigma=\sigma_{\mathrm{orig}} \cdot k$). Similarly, for the ground-based data we fit a jitter term that is added in quadrature to the Poisson noise ($\sigma^2 = \sigma_\mathrm{Poisson}^2 + \sigma_\mathrm{systematic}^2$).

Our systematics model was designed to capture variations in all the ground-based light curves due to changing atmospheric conditions and instrumental effects. This model incorporated weighted contributions from comparison star light curves, with the weights treated as free parameters. We also included weights for the airmass, PSF width, sky background flux, and distances from the median centroid position as candidate covariates. To optimize the systematics model for each ground-based light curve, we selected the combination of detrending parameters that minimized the Bayesian Information Criterion (BIC; \citealt{Schwarz1978}) prior to performing the joint fit with the phase-folded TESS data.

\section{Statistical Validation}
\label{sec:validation}

For each TESS candidate, we calculated the probability that it might be an astrophysical false positive using \texttt{TRICERATOPS+}\footnote{\url{https://github.com/JGB276/TRICERATOPS-plus}}\citep{GomezBarrientos2025}, a modified version of \texttt{TRICERATOPS} \citep{Giacalone2021}. This Python package employs a Bayesian framework that combines information about the transit shape measured by TESS and (when available) adaptive optics imaging data with prior knowledge of stellar populations in the Milky Way to calculate both the false positive probability (FPP) and nearby false positive probability (NFPP) for each planet candidate. The FPP quantifies the likelihood that a transit signal does not originate from a planet transiting the target star, while the NFPP represents the probability that the observed transit signal stems from a resolved nearby star. In \citet{GomezBarrientos2025}, we updated this framework to incorporate information from multi-color ground-based transit photometry, which has a higher angular resolution than TESS and can significantly reduce both the FPP and the NFPP. Following \citet{Giacalone2021}, we require planet candidates to satisfy FPP < 0.015 and NFPP $< 10^{-3}$ to be considered statistically validated planets. We note that for most of our targets, the FPP values are orders of magnitude below these thresholds. Among our five targets, TOI-5716~b has the highest FPP at $4.3 \times 10^{-3}$. As recommended by \citet{Giacalone2021}, we calculate the FPP and NFPP 20 times and report the mean value as the nominal value along with the 68$\%$ confidence interval.

\section{Results}
\label{sec:results}

\begin{table*}[]
\centering
\fontsize{7.5pt}{7.5pt}\selectfont
\begin{tabular}{lllllllll}
\hline
TOI & $t_0$-2457000\tablenotemark{\scriptsize a} & $P$\tablenotemark{\scriptsize b} & $a/R_*$ & $b$ & $i$ & $R_p/R_*$\tablenotemark{\scriptsize c} & $R_p$ & $T_\mathrm{eq}$\tablenotemark{\scriptsize d}  \\ 
    & (BTJD)        & (days)                                  &         &     &(deg) &          & ($R_\oplus$) & (K)
\\ \hline
5489~b & $1493.70710_{-0.00115}^{+0.00110}$ & $3.15222_{-0.00001}^{+0.00001}$ & $16.26_{-0.48}^{+0.48}$ & $0.391_{-0.133}^{+0.083}$ & $88.6 \pm 0.50$ & $0.0308_{-0.0010}^{+0.0010}$ & $1.40_{-0.05}^{+0.05}$  & 624 \\
5489~c & $1496.32542_{-0.00210}^{+0.00215}$ & $4.92126_{-0.00001}^{+0.00001}$ & $21.83_{-0.66}^{+0.66}$ & $0.449_{-0.138}^{+0.080}$ & $88.8 \pm 0.40$ & $0.0281_{-0.0014}^{+0.0014}$ & $1.28_{-0.07}^{+0.07}$ & 536 \\
5716~b & $1683.61514_{-0.00161}^{+0.00164}$ & $6.76630_{-0.00001}^{+0.00001}$ & $39.00_{-1.21}^{+1.17}$ & $0.410_{-0.102}^{+0.075}$ & $89.40 \pm 0.15$ & $0.0390_{-0.0014}^{+0.0014}$ & $0.96_{-0.05}^{+0.05}$ & 400 \\
5728~b & $1685.08679_{-0.00118}^{+0.00120}$ & $11.49761_{-0.00003}^{+0.00003}$ & $39.90_{-0.98}^{+0.78}$ & $0.117_{-0.083}^{+0.107}$ & $89.83 \pm 0.15$ & $0.0319_{-0.0009}^{+0.0009}$ & $1.31_{-0.05}^{+0.05}$ & 382 \\
5736~b & $1738.89683_{-0.00143}^{+0.00155}$ & $0.64899_{-0.00001}^{+0.00001}$ & $4.53_{-0.13}^{+0.13}$ & $0.435_{-0.242}^{+0.145}$ & $84.5 \pm 2.2$ & $0.0247_{-0.0007}^{+0.0007}$ & $1.56_{-0.07}^{+0.07}$ & 1311 \\
\hline
\end{tabular}
\caption{Posterior results from the joint fit of the TESS and ground-based light curves.}
\label{tab:posterior}
\vspace{0.1cm}
\footnotesize
\begin{flushleft}
\textbf{Notes.}
\tablenotemark{\scriptsize a} Transit mid-time of the first transit in the TESS timeseries.
\tablenotemark{\scriptsize b} Derived from a separate transit fit to the TESS data.
\tablenotemark{\scriptsize c} Weighted average of $R_p/R_*$ values from TESS and ground-based data.
\tablenotemark{\scriptsize d} The planet’s equilibrium temperature assuming full day-night heat redistribution and zero albedo.
\end{flushleft}
\end{table*}

\begin{figure*}
    \centering
    \includegraphics[width=0.95\linewidth]{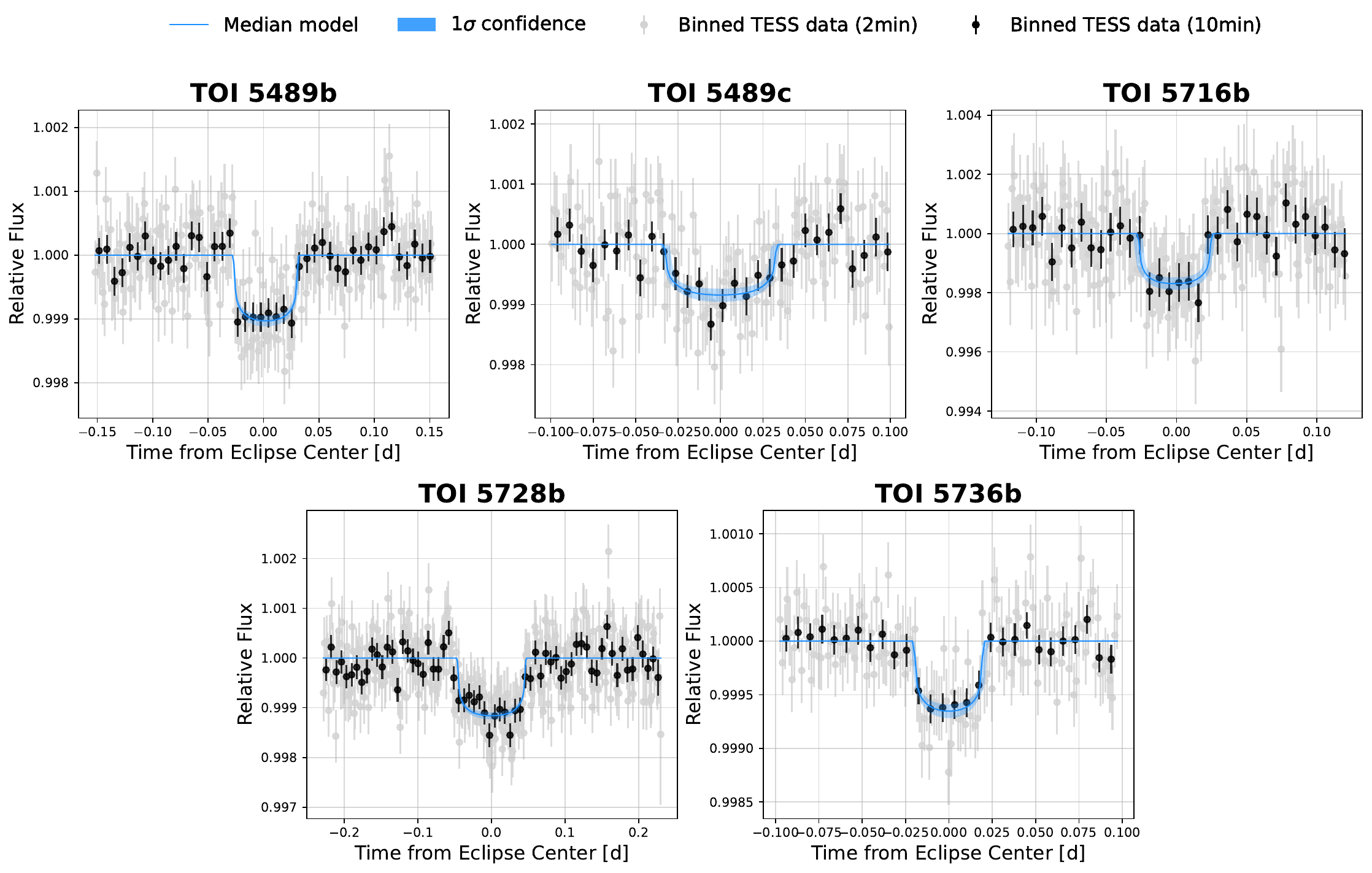}
    \caption{Stacked TESS light curves for our five targets. Grey points show the 2-minute binned data, black points show the 10-minute binned data. The blue line represents the median transit model from the posterior distribution, with light blue shading indicating the 68$\%$ confidence interval.}
    \label{fig:tess}
\end{figure*}

\begin{figure*}
    \centering
    \includegraphics[width=0.95\linewidth]{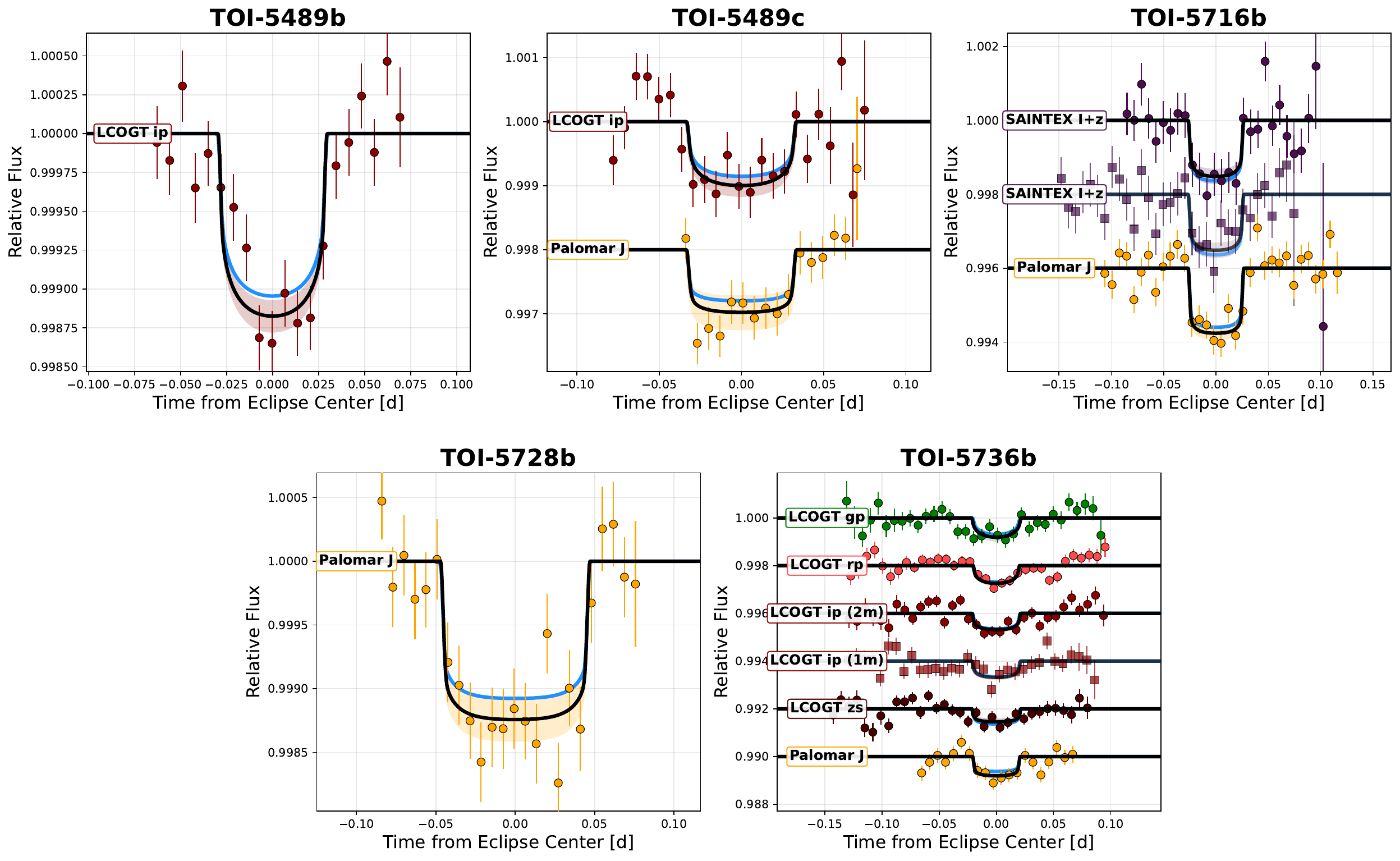}
    \caption{Multi-color ground-based light curves for our five targets. Data points represent 10-minute binned photometry, with light curves obtained using identical filters and exposure times phase-folded together. Black curves show the median transit model derived from the posterior distribution, with colored shading indicating the 68$\%$ confidence intervals. Blue curves represent the nominal TESS light curve model projected into the corresponding ground-based filters.}
    \label{fig:ground_lcs}
\end{figure*}

\subsection{TOI-5489~b}

TOI-5489.01 is a super-Earth sized ($\mathrm{R_p = 1.40}$ $\mathrm{R_\earth}$) planet candidate that orbits its M dwarf host star ($V=$ 13.8; $J=$ 10.3) every 3.15 days. It was identified simultaneously by the QLP-FAINT and SPOC TPS pipelines \citep{Huang2020, Huang2020b, Kunimoto2021, Jenkins2002, Jenkins2010, Jenkins2020}. The QLP-FAINT pipeline employs a Box Least Squares (BLS; \citet{Kovacs2002}) transit search algorithm while the SPOC transit search utilizes an adaptive, noise-compensating matched filter. Our observations indicate that TOI-5489.01 is a transiting planet. The high-resolution image from the PHARO instrument on Palomar in the $K$-continuum filter shows no nearby stellar companions (see Figure \ref{fig:ccs}). The phase-folded TESS light curve (Sectors 7 - 72) in Figure \ref{fig:tess} shows a U-shaped transit signal, while the phase-folded LCOGT $i$ band light curve in Figure \ref{fig:ground_lcs} indicates that the transit signal is slightly chromatic (inconsistent at 1.2$\sigma$ level). Our statistical validation analysis with \texttt{TRICERATOPS+} finds that FPP=$3 \times 10^{-4}$ (68$\%$ confidence interval of $2 \times 10^{-4}$ to $4 \times 10^{-4}$) and that the NFPP is zero, since there are no resolved stars inside the optimal aperture of the LCOGT light curve, based on the TESS photometry, LCOGT data, and the Palomar/PHARO contrast curve. With FPP$<$0.015 and NFPP$<10^{-3}$, TOI-5489.01 satisfies the criteria for a statistically validated planet and we hereafter refer to it as TOI-5489~b. 

\subsection{TOI-5489~c}

TOI-5489.02 (also identified by the SPOC TPS pipeline; \citet{Jenkins2002, Jenkins2010, Jenkins2020}) is a super-Earth sized ($\mathrm{R_p = 1.28}$ $\mathrm{R_\earth}$) planet candidate orbiting exterior to TOI-5489~b with a period of 4.92 days yielding a period ratio that places them within $4\%$ of the 3:2 orbital resonance. While this proximity suggests potential dynamical interactions, Kepler systems near first-order resonance with significant long-term transit timing variations (TTVs) typically show resonance proximity closer than 3$\%$ \citep[see Table 8 of][]{Holczer2016}. Given this $4\%$
separation, we adopt a linear ephemeris when stacking the TESS transits. Like TOI-5489.01, our observations indicate that TOI-5489.02 is also a transiting planet, consistent with previous studies demonstrating that candidates in multi-planet systems are much more likely to be bona fide planets \citep{Rowe2014}. The high-resolution image from Palomar/PHARO in the $K$-continuum filter shows no nearby stellar companions (Figure \ref{fig:ccs}). The phase-folded TESS light curve (Sectors 7 - 72; Figure \ref{fig:tess}) shows a U-shaped transit signal, while the phase-folded LCOGT $i$ band light curve and the Palomar $J$ band light curve indicate that the transit signal is achromatic (Figure \ref{fig:ground_lcs}). Our statistical validation analysis with \texttt{TRICERAOTPS+} incorporating the TESS photometry, LCOGT data, Palomar data, and the Palomar/PHARO contrast curve, indicates that FPP=$3 \times 10^{-4}$ (68$\%$ confidence interval of $2 \times 10^{-4}$ to $4 \times 10^{-4}$) and that the NFPP is zero, as there are no resolved stars inside the optimal apertures of the ground-based light curves. We note that \texttt{TRICERAOTPS+} does not account for planet multiplicity when estimating the FPP of a given candidate \citep[e.g.,][]{Guerrero2021} and thus the current estimate of this candidate is effectively an upper limit. Nevertheless, given these probabilities, TOI-5489.02 satisfies the criteria for a statistically validated planet and we hereafter refer to it as TOI-5489~c.

\subsection{TOI-5716~b}

TOI-5716.01 is an Earth-sized ($\mathrm{R_p = 0.96}$ $\mathrm{R_\earth}$) planet candidate orbiting a relatively faint M dwarf host star ($V=$ 15.8; $J=$ 11.6) every 6.77 days. It was first identified by the SPOC TPS pipeline \citep{Jenkins2002, Jenkins2010, Jenkins2020}. Our observations indicate that TOI-5716.01 is a transiting planet. The high-resolution image from SAI-2.5m \citep{Strakhov2023} in the $I$ filter shows no nearby stellar companions (Figure \ref{fig:ccs}). The phase-folded TESS data (Sectors 14 - 75; Figure \ref{fig:tess}) show a U-shaped transit signal. Furthermore, the SAINT-EX $I+z$ filter data and the Palomar $J$ band data indicate that the transit signal is achromatic (Figure \ref{fig:ground_lcs}). Based on the TESS photometry, SAINT-EX data, Palomar data, and the SAI-2.5m contrast curve, we find that FPP=$4.3 \times 10^{-3}$ (68$\%$ confidence interval of $4.0 \times 10^{-3}$ to $4.6 \times 10^{-3}$). Since there are no resolved stars inside the optimal apertures of the ground-based light curves, the NFPP is zero. As a result, TOI-5716.01 satisfies the criteria for a statistically validated planet and we hereafter refer to it as TOI-5716~b.

\subsection{TOI-5728~b}

TOI-5728.01 (identified by the SPOC TPS pipeline; \citet{Jenkins2002, Jenkins2010, Jenkins2020}) is a super-Earth sized ($\mathrm{R_p = 1.31}$ $\mathrm{R_\earth}$) planet candidate around a similarly faint M dwarf star ($V=$ 15.0; $J=$ 11.0). It orbits its host star every 11.50 days. Our observations indicate that TOI-5728.01 is a transiting planet. The high-resolution image from the NIRC2 instrument on Keck II in the $K$-continuum filter shows no nearby stellar companions (Figure \ref{fig:ccs}). The phase-folded TESS data (Sectors 14 - 86; Figure \ref{fig:tess}) show a U-shaped transit signal, while the Palomar $J$ band data confirm that the transit signal is achromatic ($R_p/R_*$ from TESS and Palomar are consistent at 1$\sigma$; Figure \ref{fig:ground_lcs}). Based on the TESS photometry, Palomar data, and the Keck2/NIRC2 contrast curve, we find that FPP=$2.3 \times 10^{-7}$ (68$\%$ confidence interval of $3.6 \times 10^{-8}$ to $1.5 \times 10^{-6}$) and that the NFPP is zero, since there are no resolved stars inside the optimal aperture of the Palomar light curve. Consequently, TOI-5728.01 satisfies the criteria for a statistically validated planet and we hereafter refer to it as TOI-5728~b.

\subsection{TOI-5736~c}

TOI-5736.01 (identified by the SPOC TPS pipeline; \citet{Jenkins2002, Jenkins2010, Jenkins2020}) is a super-Earth sized ($\mathrm{R_p = 1.56}$ $\mathrm{R_\earth}$) planet candidate around an early M dwarf star ($V=$ 13.6; $J=$ 10.8). It orbits its host star every 0.65 days. Our observations indicate that TOI-5736.01 is a transiting planet. The high-resolution image from the NIRC2 instrument on Keck II in the $K$-continuum filter shows no nearby stellar companions (Figure \ref{fig:ccs}). The phase-folded TESS data (Sectors 16 - 77; Figure \ref{fig:tess}) show a U-shaped transit signal, while the LCOGT $g$, $r$, $i$, and $z$ data and Palomar $J$ band data show that the transit signal is achromatic (Figure \ref{fig:ground_lcs}). With the TESS data, ground-based data, and the Keck2/NIRC2 contrast curve, we find that FPP=$7 \times 10^{-5}$ (68$\%$ confidence interval of $5 \times 10^{-5}$ to $9 \times 10^{-5}$). Given that there are no resolved stars inside the apertures of the ground-based light curves, we find that the NFPP is zero. Hence, TOI-5736.01 satisfies the criteria for a statistically validated planet and we hereafter refer to it as TOI-5736~b.

\section{Discussion}
\label{sec:discussion}

Having validated the planetary nature of our targets, in this section we explore how they compare to the broader population of small planets around low-mass stars and discuss key properties.

\begin{figure}
    \centering
    \includegraphics[width=0.95\columnwidth]{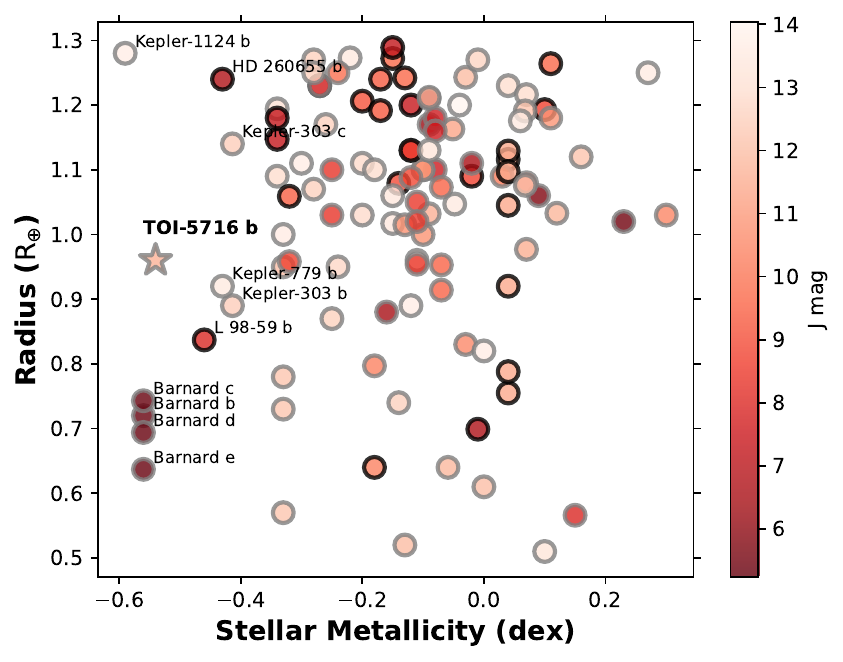}
    \caption{The population of small exoplanets around M dwarf stars ($T_*$< 4000~K) depicting the distribution of stellar metallicities for planet hosting stars. Planets with published mass measurements $>3\sigma$ are highlighted in black whereas those without are highlighted in gray. The data points are colored based on their brightness in the $J$ band as a guide for their accessibility to radial velocity measurements.}
    \label{fig:rad_met}
\end{figure}

\begin{figure}
    \centering
    \includegraphics[width=0.95\columnwidth]{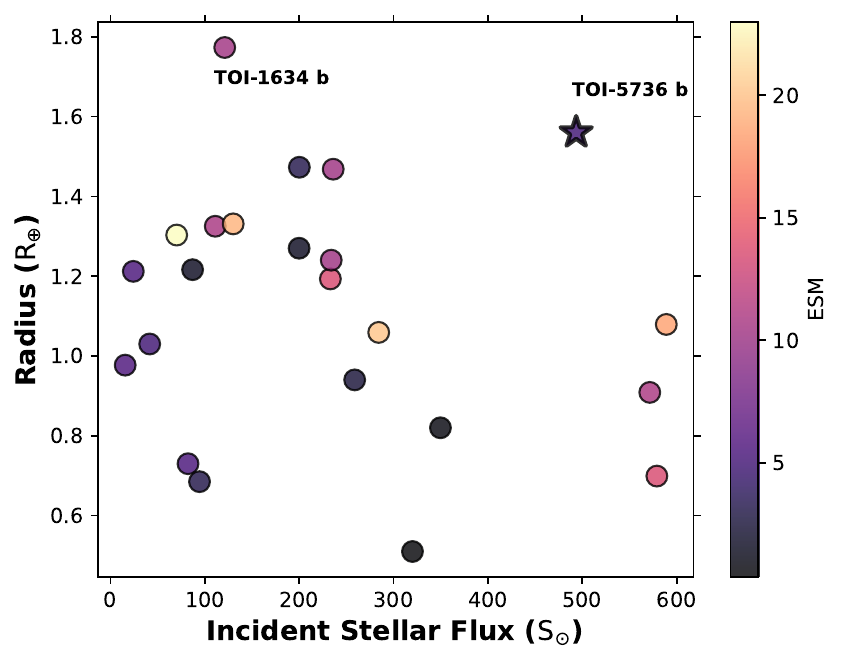}
    \caption{The population of ultra short period (\textit{P} $<$ 1 day) planets smaller than $1.8~\mathrm{R_\earth}$ around M stars. The planets are colored based on their Emission Spectroscopy Metric (ESM). TOI-5736~b has a predicted equilibrium temperature of 1311~K and an ESM of 4.}
    \label{fig:usps}
\end{figure}

\begin{figure*}
    \centering
    \includegraphics[width=0.99\textwidth]{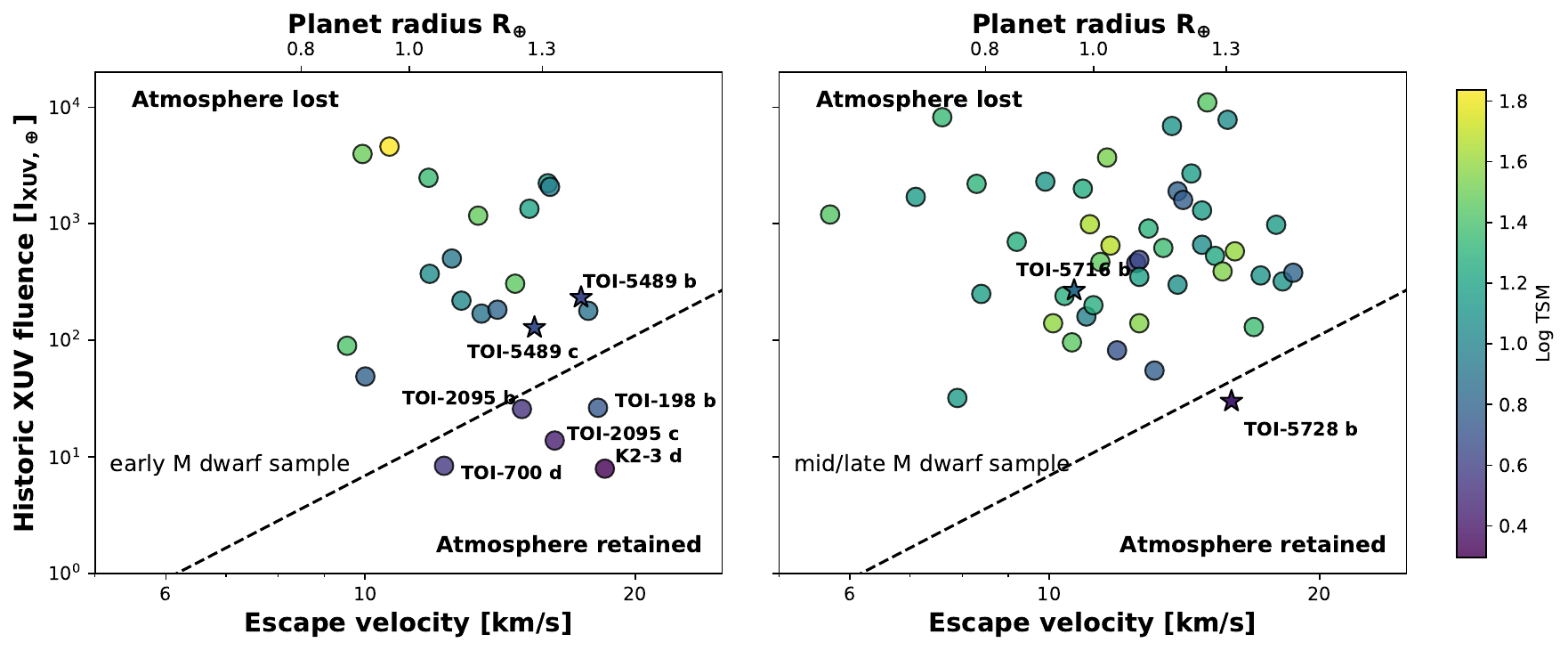}
    \caption{\textit{Left:} The cosmic shoreline for early M dwarfs (0.36 $\mathrm{M_{\odot}}$< $M_\star$< 0.6 $\mathrm{M_{\odot}}$) from \citet{Zahnle2017}. \textit{Right:} The cosmic shoreline for mid- to late-M dwarfs, as defined by \citet{Pass2025}. In both panels the circles show the population of nearby ($<$50 pc) exoplanets with radii smaller than $1.5~\mathrm{R_\earth}$ from the NASA Exoplanet Archive \citep[accessed July 15 2025;][]{Akeson2013, ps, Christiansen2025}. Our targets are shown as stars. The color of the data points correspond to their transmission spectroscopy metric.  If confirmed to have an Earth-like bulk composition, TOI-5728~b would join the very small sample of rocky planets orbiting mid- to late-M dwarfs that are expected to have retained atmospheres.}
    \label{fig:shoreline}
\end{figure*}

\subsection{Could TOI-5716 be a thick disk star?}

To date, only a small number of confirmed exoplanets have been discovered orbiting thick disk stars \citep[e.g.,][]{Campante2015, Gan2020, Weiss2021, Scott2025}. Despite the small size of this sample, comparative studies of planetary systems around stars from the thick versus thin disk have already begun to provide valuable insights into how stellar metallicity and age affect the formation and subsequent evolution of planetary systems. For instance, studies of exoplanets orbiting thick disk stars have shown that Earth-sized planets have been forming throughout most of cosmic history, providing evidence for early planet formation in metal-poor environments \citep{Campante2015}. Furthermore, observations suggest that denser rocky planets are preferentially found around younger, more metal-rich stars \citep{Weeks2025}, though this claim has been disputed by subsequent analysis \citep{Ross2025}. Lastly, occurrence rate studies have found that there is a notable deficit in planets around thick disk stars compared to the thin disk population \citep{Bashi2022, Zink2023, Hallatt2025}, suggesting that stellar metallicity and age play crucial roles in shaping planetary system architectures. 

TOI-5716 has by far the lowest metallicity of all of the stars in our sample, and is only one of a handful of known planet-hosting stars with metallicities $<-0.5$ (see Figure \ref{fig:rad_met}).  In light of this low metallicity, we investigated the possibility that this star might belong to the thick disk population by applying the probabilistic framework developed by \citet{Bensby2014}. Our analysis yielded a thick disk to thin disk probability ratio of 0.03, strongly favoring a thin disk classification. As an additional check, we performed an orbital integration using the star's Gaia-derived proper motion and radial velocity measurements. We found that the star reaches a maximum vertical height of 1.2 kpc above the Galactic plane, which is higher than the 300-400 pc scale height of the galactic thin disk. However, the eccentricity of the stellar orbit is low (e = 0.046), which is consistent with membership in the thin disk population. Our detailed abundance analysis of the alpha elements (O, Mg, Ca) further supports the classification of the star as a thin-disk member. The star exhibits only mild $\alpha$-enhancement, with a mean [$\alpha$/Fe] ratio of $+0.07 \pm 0.08$ dex, consistent with enrichment from Type II and Type Ia supernovae, as expected for chemically evolved stars in the Galactic thin disk.

Although TOI-5716 belongs to the thin disk population, its metal-poor composition still has important implications for the expected bulk composition of its Earth-sized transiting planet. Given the proposed correlation between planetary core mass fraction and host star metallicity for rocky exoplanets \citep{Brinkman2024, Adibekyan2021}, we anticipate that TOI-5716~b may have a lower core mass fraction than planets orbiting more metal-rich stars. This could be tested by obtaining a radial velocity mass measurement for this planet. To assess the feasibility of this measurement, we first calculated the expected radial velocity semi-amplitude assuming an Earth-like bulk composition based on the mass-radius relationship from \citet{Chen2017}, and found a predicted amplitude of \SI{0.97}{\meter\per\second}. As shown in Figure \ref{fig:rad_met}, there are currently only two planets smaller than 1.3~$\mathrm{R_\earth}$ that have a published mass measurement orbiting an M dwarf star with a metallicity less than $-0.4$. HD 260655~b's mass was measured using radial velocities \citep{Luque2022}, but it orbits a much brighter host star ($J$=6.67 versus 11.57 for TOI-5716) and has a larger radial velocity semi-amplitude of \SI{1.69}{\meter\per\second}. L 98-59~b's mass was also measured using radial velocities \citep{Cadieux2025, Demangeon2021, Cloutier2019}, and although it has a smaller radial velocity semi-amplitude of \SI{0.51}{\meter\per\second}, it also orbits a brighter star ($J$=7.93 versus 11.57 for TOI-5716). We conclude that it will be challenging to measure TOI-5716~b's mass with current radial velocity instruments.

Studies have also indicated that metallicity can have a strong impact on the rotational evolution of low-mass stars \citep[e.g.,][]{Amard2020, See2024, See2025}. For a fixed mass and age, metal-poor stars rotate faster than their solar metallicity counterparts but have a higher stellar Rossby number, which is inversely proportional to stellar activity \citep{Amard2020}. Therefore, TOI-5716 may exhibit lower stellar activity than more metal-rich stars, creating a more favorable stellar environment for TOI-5716~b and potentially enhancing the prospects for atmospheric retention.

\subsection{Location of newly confirmed super-Earth planets relative to the cosmic shoreline}

Most M dwarf planets smaller than 1.5~$\mathrm{R_\earth}$ with measured masses appear to have densities consistent with Earth-like bulk compositions \citep{Luque2022}. If these planets have atmospheres they must be comprised of high mean molecular weight gases, and do not contribute significantly to the planet's measured radius.  However, many close-in rocky planets orbiting M stars appear to have had their outgassed atmospheres stripped away by XUV irradiation, as discussed in \S\ref{sec:intro}. Here, we assess the likelihood that the planets in our sample with radii less than 1.5~R$_\earth$ have retained high mean molecular weight atmospheres, and examine the feasibility of detecting atmospheric absorption features using transmission spectroscopy. 

In Figure \ref{fig:shoreline}, we show the population of nearby M dwarf exoplanets ($<$ 50 pc) smaller than 1.5~$\mathrm{R_\earth}$ as a function of historic XUV fluence and escape velocity.  We plot the sample of planets orbiting early M dwarfs and mid- to late-M dwarfs separately, and calculate the XUV fluence and cosmic shoreline location following \citet{Zahnle2017} for the early M dwarf sample and \cite{Pass2025} for the mid-to-late M dwarf sample. For planets without a mass measurement, including the four new planets reported here, we assume an Earth-like bulk density and adopt the mass-radius relationship presented in \citet{Chen2017} to calculate their escape velocities. As shown in Figure \ref{fig:shoreline}, TOI-5489~b and TOI-5489~c lie above the predicted cosmic shoreline and therefore may have lost their atmospheres; however, their proximity to the shoreline means that observational limits on their atmospheric masses would help to constrain the location and slope of this boundary. For mid- to late-M dwarfs, Figure \ref{fig:shoreline} shows that almost all currently known targets lie above the cosmic shoreline, including TOI-5716~b. However, TOI-5728~b lies below the shoreline, suggesting that it is more likely to have retained its atmosphere. This is consistent with findings from \citet{Pass2025}, who note that only the largest terrestrial planets around mid- to late-M dwarfs are expected to retain their atmospheres. TOI-5728 b has a predicted radial velocity semi-amplitude of \SI{1.5}{\meter\per\second} and orbits an optically faint star ($V=15.83$, $J=11.57)$, suggesting that it will be challenging to confirm whether or not this planet has an Earth-like bulk density via a radial velocity mass measurement.

Next, we explore the detectability of atmospheric absorption features during the transit using the Transmission Spectroscopy Metric (TSM; \citet{Kempton2018}). As shown in Figure \ref{fig:shoreline}, the four planets in our sample have TSM values below the typical threshold value of 12 for planets with $\mathrm{R_{p}}$ < $1.5~\mathrm{R_\earth}$, which is generally considered to be a lower bound for atmospheric characterization studies with JWST \citep{Kempton2018}. Although this suggests that it would require a significant investment of telescope time to obtain a useful transmission spectrum, this may be justified by TOI-5728~b's unique position as the only validated planet that lies below the cosmic shoreline for mid- to late-M dwarfs.

\subsection{Does TOI-5736~b have a volatile-rich envelope?}

TOI-5736~b has the highest equilibrium temperature (1311~K) and largest radius ($\mathrm{1.56\pm0.07}$ $\mathrm{R_\earth}$) of the planets in our sample. If we compare TOI-5736~b to the sample of M dwarf planets with measured masses and radii \citep[e.g.,][]{Luque2022, Wanderley2025}, we find that it is on the upper end of the range for planets with Earth-like bulk compositions. Alternatively, it could have retained a volatile-rich envelope despite its relatively high XUV irradiation level. TOI-1634~b \citep[e.g.,][]{Cloutier2021} is similar in size and also orbits an M dwarf on a sub-day orbital period (see Figure \ref{fig:usps}). Radial velocity observations of this planet indicate that it has retained a volatile-rich atmosphere, but it also receives a bolometric flux that is approximately a factor of three less than that of TOI-5736~b. 

If the mass of TOI-5736~b could be measured using the radial velocity technique, it would be straightforward to determine whether or not it also hosts a volatile-rich envelope. Assuming an Earth-like bulk density based on the mass-radius relationship presented in \citet{Chen2017}, TOI-5736~b is expected to have a radial velocity semi-amplitude of \SI{5.5}{\meter\per\second}. This is comparable to the measured radial velocity semi-amplitude of TOI-1634~b \citep[$5.1\pm$\SI{0.8}{\meter\per\second};][]{Cloutier2021} and the two stars are similarly bright ($V=13.55$, $J=10.83$ for TOI~5736; $V=13.22$, $J=9.48$ for TOI-1634).  This suggests that it should be straightforward to measure TOI~5736~b's mass with current radial velocity instruments.  Alternatively, the presence or absence of a thick atmosphere can also be constrained by measuring thermal emission from the planet's dayside.  We quantify the relative favorability of TOI~5736~b for this measurement using the Emission Spectroscopy Metric (ESM; \citet{Kempton2018}), as indicated by the colored shading in Fig. \ref{fig:usps}.

\section{Conclusions}
\label{sec:conc}

In this paper, we statistically validate five small transiting planets around M-dwarfs using 2-minute TESS light curves, multi-color ground-based light curves, and high-resolution imaging. The sample consists of TOI-5489~b, TOI-5489~c, TOI-5716~b, TOI-5728~b, and TOI-5736~b. These five planets span a range of orbital periods, from TOI-5736~b, which orbits its host every 15.6 hours, to TOI-5728~b, which has an orbital period of 11.50 days. Although the targets in this sample will be challenging for atmospheric studies compared to other similar systems, these systems represent an important addition to the sample of planets that could be used to test the cosmic shoreline framework. If radial velocity measurements confirm an Earth-like density, TOI-5728~b would join the population of small planets around mid-to-late M stars that lie below the cosmic shoreline.

\section{Acknowledgments}

We thank the anonymous referee for a helpful report. We would like to thank the Palomar Observatory staff and telescope operators, including Kathleen Koviak, Paul Nied, Diana Roderick, Rigel Rafto, and John Stone for their assistance with the observations of TOI-5489, TOI-5716, TOI-5728, and TOI-5736. This work is based on observations obtained at the Hale Telescope, Palomar Observatory, as part of a collaborative agreement between the Caltech Optical Observatories and the Jet Propulsion Laboratory (operated by Caltech for NASA). Some of the observations presented in this paper were carried out at the Observatorio Astron\'omico Nacional on the Sierra de San Pedro M\'artir (OAN-SPM), Baja California, M\'exico.

This research was carried out at the Jet Propulsion Laboratory and the California Institute of Technology under a contract with the National Aeronautics and Space Administration and funded through the President’s and Director’s Research \& Development Fund Program. We also acknowledge support from the Swiss National Science Foundation SPIRIT-216537 and the Centre for Space and Habitability (CSH) of the University of Bern.
Part of this work received support from the National Centre for Competence in Research PlanetS, supported by the Swiss National Science Foundation (SNSF). 
YGMC and AK are partially supported by UNAM PAPIIT-IG101224.
B.-O. D. acknowledges support from the Swiss State Secretariat for Education, Research and Innovation (SERI) under contract number MB22.00046.

This paper includes data collected by the TESS mission that are publicly available from the Mikulski Archive for Space Telescopes (MAST). The specific observations analyzed can be accessed via \dataset[doi:10.17909/c3pd-m382]{https://doi.org/10.17909/c3pd-m382}. Funding for the TESS mission is provided by NASA's Science Mission Directorate. We acknowledge the use of public TESS data from pipelines at the TESS Science Office and at the TESS Science Processing Operations Center. Resources supporting this work were provided by the NASA High-End Computing (HEC) Program through the NASA Advanced Supercomputing (NAS) Division at Ames Research Center for the production of the SPOC data products. This research has made use of the Exoplanet Follow-up Observation Program website, which is operated by the California Institute of Technology, under contract with the National Aeronautics and Space Administration under the Exoplanet Exploration Program.

We acknowledge financial support from the Agencia Estatal de Investigaci\'on of the Ministerio de Ciencia e Innovaci\'on MCIN/AEI/10.13039/501100011033 and the ERDF “A way of making Europe” through project PID2021-125627OB-C32, and from the Centre of Excellence “Severo Ochoa” award to the Instituto de Astrofisica de Canarias.

This work makes use of observations from the Las Cumbres Observatory global telescope network.

This material is based upon work supported by the National Science Foundation Graduate Research Fellowship Program under Grant No. DGE‐1745301.

M.G. is FNRS Research Director

F. M. acknowledges the financial support from the Agencia Estatal de Investigaci\'{o}n del Ministerio de Ciencia, Innovaci\'{o}n y Universidades (MCIU/AEI) through grant PID2023-152906NA-I00.

WGL gratefully acknowledges support from the Department of Defense's National Defense Science \& Engineering Graduate (NDSEG) Fellowship. WGL also thanks the LSST-DA Data Science Fellowship Program, which is funded by LSST-DA, the Brinson Foundation, the WoodNext Foundation, and the Research Corporation for Science Advancement Foundation; his participation in the program has benefited this work.

ML acknowledges support of the Swiss National Science Foundation under grant number PCEFP2194576. The contribution of ML has been carried out within the framework of the NCCR PlanetS supported by the Swiss National Science Foundation under grant 51NF40205606.

This work is partly supported by JSPS KAKENHI Grant Number JP24H00017 and JSPS Bilateral Program Number JPJSBP120249910. This paper is based on observations made with the MuSCAT instruments, developed by the Astrobiology Center (ABC) in Japan, the University of Tokyo, and Las Cumbres Observatory (LCOGT). MuSCAT3 was developed with financial support by JSPS KAKENHI (JP18H05439) and JST PRESTO (JPMJPR1775), and is located at the Faulkes Telescope North on Maui, HI (USA), operated by LCOGT.

Y. G. M. C. and B.-O. D. acknowledges support from the Swiss National Science Foundation (IZSTZ0216537).

D.S. acknowledges support from the Foundation for Research and Technological Innovation Support of the State of Sergipe (FAPITEC/SE) and the National Council for Scientific and Technological Development (CNPq), under grant numbers 794017/2013 and 444372/2024-5.

DRC acknowledges partial support from NASA Grant 182XRP182-0007.

Some of the data presented herein were obtained at Keck Observatory, which is a private 501(c)3 non-profit organization operated as a scientific partnership among the California Institute of Technology, the University of California, and the National Aeronautics and Space Administration. The Observatory was made possible by the generous financial support of the W. M. Keck Foundation.

The authors wish to recognize and acknowledge the very significant cultural role and reverence that the summit of Maunakea has always had within the Native Hawaiian community. We are most fortunate to have the opportunity to conduct observations from this mountain.

\bibliography{main}{}

\begin{thebibliography}{}
\expandafter\ifx\csname natexlab\endcsname\relax\def\natexlab#1{#1}\fi
\providecommand{\url}[1]{\href{#1}{#1}}
\providecommand{\dodoi}[1]{doi:~\href{http://doi.org/#1}{\nolinkurl{#1}}}
\providecommand{\doeprint}[1]{\href{http://ascl.net/#1}{\nolinkurl{http://ascl.net/#1}}}
\providecommand{\doarXiv}[1]{\href{https://arxiv.org/abs/#1}{\nolinkurl{https://arxiv.org/abs/#1}}}

\bibitem[{{Adibekyan} {et~al.}(2021){Adibekyan}, {Dorn}, {Sousa}, {Santos}, {Bitsch}, {Israelian}, {Mordasini}, {Barros}, {Delgado Mena}, {Demangeon}, {Faria}, {Figueira}, {Hakobyan}, {Oshagh}, {Soares}, {Kunitomo}, {Takeda}, {Jofr{\'e}}, {Petrucci}, \& {Martioli}}]{Adibekyan2021}
{Adibekyan}, V., {Dorn}, C., {Sousa}, S.~G., {et~al.} 2021, Science, 374, 330, \dodoi{10.1126/science.abg8794}

\bibitem[{{Ahumada} {et~al.}(2020){Ahumada}, {Allende Prieto}, {Almeida}, {Anders}, {Anderson}, {Andrews}, {Anguiano}, {Arcodia}, {Armengaud}, {Aubert}, {Avila}, {Avila-Reese}, {Badenes}, {Balland}, {Barger}, {Barrera-Ballesteros}, {Basu}, {Bautista}, {Beaton}, {Beers}, {Benavides}, {Bender}, {Bernardi}, {Bershady}, {Beutler}, {Bidin}, {Bird}, {Bizyaev}, {Blanc}, {Blanton}, {Boquien}, {Borissova}, {Bovy}, {Brandt}, {Brinkmann}, {Brownstein}, {Bundy}, {Bureau}, {Burgasser}, {Burtin}, {Cano-D{\'\i}az}, {Capasso}, {Cappellari}, {Carrera}, {Chabanier}, {Chaplin}, {Chapman}, {Cherinka}, {Chiappini}, {Doohyun Choi}, {Chojnowski}, {Chung}, {Clerc}, {Coffey}, {Comerford}, {Comparat}, {da Costa}, {Cousinou}, {Covey}, {Crane}, {Cunha}, {Ilha}, {Dai}, {Damsted}, {Darling}, {Davidson}, {Davies}, {Dawson}, {De}, {de la Macorra}, {De Lee}, {Queiroz}, {Deconto Machado}, {de la Torre}, {Dell'Agli}, {du Mas des Bourboux}, {Diamond-Stanic}, {Dillon}, {Donor}, {Drory}, {Duckworth}, {Dwelly}, {Ebelke}, {Eftekharzadeh}, {Davis
  Eigenbrot}, {Elsworth}, {Eracleous}, {Erfanianfar}, {Escoffier}, {Fan}, {Farr}, {Fern{\'a}ndez-Trincado}, {Feuillet}, {Finoguenov}, {Fofie}, {Fraser-McKelvie}, {Frinchaboy}, {Fromenteau}, {Fu}, {Galbany}, {Garcia}, {Garc{\'\i}a-Hern{\'a}ndez}, {Garma Oehmichen}, {Ge}, {Geimba Maia}, {Geisler}, {Gelfand}, {Goddy}, {Gonzalez-Perez}, {Grabowski}, {Green}, {Grier}, {Guo}, {Guy}, {Harding}, {Hasselquist}, {Hawken}, {Hayes}, {Hearty}, {Hekker}, {Hogg}, {Holtzman}, {Horta}, {Hou}, {Hsieh}, {Huber}, {Hunt}, {Ider Chitham}, {Imig}, {Jaber}, {Jimenez Angel}, {Johnson}, {Jones}, {J{\"o}nsson}, {Jullo}, {Kim}, {Kinemuchi}, {Kirkpatrick}, {Kite}, {Klaene}, {Kneib}, {Kollmeier}, {Kong}, {Kounkel}, {Krishnarao}, {Lacerna}, {Lan}, {Lane}, {Law}, {Le Goff}, {Leung}, {Lewis}, {Li}, {Lian}, {Lin}, {Long}, {Longa-Pe{\~n}a}, {Lundgren}, {Lyke}, {Mackereth}, {MacLeod}, {Majewski}, {Manchado}, {Maraston}, {Martini}, {Masseron}, {Masters}, {Mathur}, {McDermid}, {Merloni}, {Merrifield}, {M{\'e}sz{\'a}ros}, {Miglio}, {Minniti},
  {Minsley}, {Miyaji}, {Mohammad}, {Mosser}, {Mueller}, {Muna}, {Mu{\~n}oz-Guti{\'e}rrez}, {Myers}, {Nadathur}, {Nair}, {Nandra}, {Correa do Nascimento}, {Nevin}, {Newman}, {Nidever}, {Nitschelm}, {Noterdaeme}, {O'Connell}, {Olmstead}, {Oravetz}, {Oravetz}, {Osorio}, {Pace}, {Padilla}, {Palanque-Delabrouille}, \& {Palicio}}]{Ahumada2020}
{Ahumada}, R., {Allende Prieto}, C., {Almeida}, A., {et~al.} 2020, \apjs, 249, 3, \dodoi{10.3847/1538-4365/ab929e}

\bibitem[{{Akeson} {et~al.}(2013){Akeson}, {Chen}, {Ciardi}, {Crane}, {Good}, {Harbut}, {Jackson}, {Kane}, {Laity}, {Leifer}, {Lynn}, {McElroy}, {Papin}, {Plavchan}, {Ram{\'\i}rez}, {Rey}, {von Braun}, {Wittman}, {Abajian}, {Ali}, {Beichman}, {Beekley}, {Berriman}, {Berukoff}, {Bryden}, {Chan}, {Groom}, {Lau}, {Payne}, {Regelson}, {Saucedo}, {Schmitz}, {Stauffer}, {Wyatt}, \& {Zhang}}]{Akeson2013}
{Akeson}, R.~L., {Chen}, X., {Ciardi}, D., {et~al.} 2013, \pasp, 125, 989, \dodoi{10.1086/672273}

\bibitem[{Allard {et~al.}(2000)Allard, Hauschildt, Alexander, Tamanai, \& Schweitzer}]{allard2000}
Allard, F., Hauschildt, P.~H., Alexander, D.~R., Tamanai, A., \& Schweitzer, A. 2000, The Astrophysical Journal, 540, 1005, \dodoi{10.1086/309290}

\bibitem[{{Amard} \& {Matt}(2020)}]{Amard2020}
{Amard}, L., \& {Matt}, S.~P. 2020, \apj, 889, 108, \dodoi{10.3847/1538-4357/ab6173}

\bibitem[{{Bashi} \& {Zucker}(2022)}]{Bashi2022}
{Bashi}, D., \& {Zucker}, S. 2022, \mnras, 510, 3449, \dodoi{10.1093/mnras/stab3596}

\bibitem[{{Bensby} {et~al.}(2014){Bensby}, {Feltzing}, \& {Oey}}]{Bensby2014}
{Bensby}, T., {Feltzing}, S., \& {Oey}, M.~S. 2014, \aap, 562, A71, \dodoi{10.1051/0004-6361/201322631}

\bibitem[{{Berger} {et~al.}(2020){Berger}, {Huber}, {van Saders}, {Gaidos}, {Tayar}, \& {Kraus}}]{Berger2020}
{Berger}, T.~A., {Huber}, D., {van Saders}, J.~L., {et~al.} 2020, \aj, 159, 280, \dodoi{10.3847/1538-3881/159/6/280}

\bibitem[{{Berger} {et~al.}(2023){Berger}, {Schlieder}, \& {Huber}}]{Berger2023}
{Berger}, T.~A., {Schlieder}, J.~E., \& {Huber}, D. 2023, arXiv e-prints, arXiv:2301.11338, \dodoi{10.48550/arXiv.2301.11338}

\bibitem[{{Birky} {et~al.}(2020){Birky}, {Hogg}, {Mann}, \& {Burgasser}}]{Birky2020}
{Birky}, J., {Hogg}, D.~W., {Mann}, A.~W., \& {Burgasser}, A. 2020, \apj, 892, 31, \dodoi{10.3847/1538-4357/ab7004}

\bibitem[{{Brinkman} {et~al.}(2024){Brinkman}, {Polanski}, {Huber}, {Weiss}, {Valencia}, \& {Plotnykov}}]{Brinkman2024}
{Brinkman}, C.~L., {Polanski}, A.~S., {Huber}, D., {et~al.} 2024, \aj, 168, 281, \dodoi{10.3847/1538-3881/ad82eb}

\bibitem[{{Brown} {et~al.}(2013){Brown}, {Baliber}, {Bianco}, {Bowman}, {Burleson}, {Conway}, {Crellin}, {Depagne}, {De Vera}, {Dilday}, {Dragomir}, {Dubberley}, {Eastman}, {Elphick}, {Falarski}, {Foale}, {Ford}, {Fulton}, {Garza}, {Gomez}, {Graham}, {Greene}, {Haldeman}, {Hawkins}, {Haworth}, {Haynes}, {Hidas}, {Hjelstrom}, {Howell}, {Hygelund}, {Lister}, {Lobdill}, {Martinez}, {Mullins}, {Norbury}, {Parrent}, {Paulson}, {Petry}, {Pickles}, {Posner}, {Rosing}, {Ross}, {Sand}, {Saunders}, {Shobbrook}, {Shporer}, {Street}, {Thomas}, {Tsapras}, {Tufts}, {Valenti}, {Vander Horst}, {Walker}, {White}, \& {Willis}}]{Brown:2013}
{Brown}, T.~M., {Baliber}, N., {Bianco}, F.~B., {et~al.} 2013, \pasp, 125, 1031, \dodoi{10.1086/673168}

\bibitem[{{Cadieux} {et~al.}(2025){Cadieux}, {L'Heureux}, {Piaulet-Ghorayeb}, {Doyon}, {Artigau}, {Cook}, {Coulombe}, {Roy}, {Lafreni{\`e}re}, {Lamontagne}, {Radica}, {Benneke}, {Ahrer}, {Weisserman}, \& {Cloutier}}]{Cadieux2025}
{Cadieux}, C., {L'Heureux}, A., {Piaulet-Ghorayeb}, C., {et~al.} 2025, \aj, 170, 154, \dodoi{10.3847/1538-3881/adef59}

\bibitem[{{Campante} {et~al.}(2015){Campante}, {Barclay}, {Swift}, {Huber}, {Adibekyan}, {Cochran}, {Burke}, {Isaacson}, {Quintana}, {Davies}, {Silva Aguirre}, {Ragozzine}, {Riddle}, {Baranec}, {Basu}, {Chaplin}, {Christensen-Dalsgaard}, {Metcalfe}, {Bedding}, {Handberg}, {Stello}, {Brewer}, {Hekker}, {Karoff}, {Kolbl}, {Law}, {Lundkvist}, {Miglio}, {Rowe}, {Santos}, {Van Laerhoven}, {Arentoft}, {Elsworth}, {Fischer}, {Kawaler}, {Kjeldsen}, {Lund}, {Marcy}, {Sousa}, {Sozzetti}, \& {White}}]{Campante2015}
{Campante}, T.~L., {Barclay}, T., {Swift}, J.~J., {et~al.} 2015, \apj, 799, 170, \dodoi{10.1088/0004-637X/799/2/170}

\bibitem[{{Casey} {et~al.}(2016){Casey}, {Hogg}, {Ness}, {Rix}, {Ho}, \& {Gilmore}}]{Casey2016}
{Casey}, A.~R., {Hogg}, D.~W., {Ness}, M., {et~al.} 2016, arXiv e-prints, arXiv:1603.03040, \dodoi{10.48550/arXiv.1603.03040}

\bibitem[{{Chen} \& {Kipping}(2017)}]{Chen2017}
{Chen}, J., \& {Kipping}, D. 2017, \apj, 834, 17, \dodoi{10.3847/1538-4357/834/1/17}

\bibitem[{{Choi} {et~al.}(2016){Choi}, {Dotter}, {Conroy}, {Cantiello}, {Paxton}, \& {Johnson}}]{Choi2016}
{Choi}, J., {Dotter}, A., {Conroy}, C., {et~al.} 2016, \apj, 823, 102, \dodoi{10.3847/0004-637X/823/2/102}

\bibitem[{{Christiansen} {et~al.}(2025){Christiansen}, {McElroy}, {Harbut}, {Ciardi}, {Crane}, {Good}, {Hardegree-Ullman}, {Kesseli}, {Lund}, {Lynn}, {Muthiar}, {Nilsson}, {Oluyide}, {Papin}, {Rivera}, {Swain}, {Susemiehl}, {Tam}, {van Eyken}, \& {Beichman}}]{Christiansen2025}
{Christiansen}, J.~L., {McElroy}, D.~L., {Harbut}, M., {et~al.} 2025, arXiv e-prints, arXiv:2506.03299, \dodoi{10.48550/arXiv.2506.03299}

\bibitem[{{Cloutier} {et~al.}(2019){Cloutier}, {Astudillo-Defru}, {Bonfils}, {Jenkins}, {Berdi{\~n}as}, {Ricker}, {Vanderspek}, {Latham}, {Seager}, {Winn}, {Jenkins}, {Almenara}, {Bouchy}, {Delfosse}, {D{\'\i}az}, {D{\'\i}az}, {Doyon}, {Figueira}, {Forveille}, {Kurtovic}, {Lovis}, {Mayor}, {Menou}, {Morgan}, {Morris}, {Muirhead}, {Murgas}, {Pepe}, {Santos}, {S{\'e}gransan}, {Smith}, {Tenenbaum}, {Torres}, {Udry}, {Vezie}, \& {Villasenor}}]{Cloutier2019}
{Cloutier}, R., {Astudillo-Defru}, N., {Bonfils}, X., {et~al.} 2019, \aap, 629, A111, \dodoi{10.1051/0004-6361/201935957}

\bibitem[{{Cloutier} {et~al.}(2021){Cloutier}, {Charbonneau}, {Stassun}, {Murgas}, {Mortier}, {Massey}, {Lissauer}, {Latham}, {Irwin}, {Haywood}, {Guerra}, {Girardin}, {Giacalone}, {Bosch-Cabot}, {Bieryla}, {Winn}, {Watson}, {Vanderspek}, {Udry}, {Tamura}, {Sozzetti}, {Shporer}, {S{\'e}gransan}, {Seager}, {Savel}, {Sasselov}, {Rose}, {Ricker}, {Rice}, {Quintana}, {Quinn}, {Piotto}, {Phillips}, {Pepe}, {Pedani}, {Parviainen}, {Palle}, {Narita}, {Molinari}, {Micela}, {McDermott}, {Mayor}, {Matson}, {Martinez Fiorenzano}, {Lovis}, {L{\'o}pez-Morales}, {Kusakabe}, {Jensen}, {Jenkins}, {Huang}, {Howell}, {Harutyunyan}, {F{\H{u}}r{\'e}sz}, {Fukui}, {Esquerdo}, {Esparza-Borges}, {Dumusque}, {Dressing}, {Fabrizio}, {Collins}, {Cameron}, {Christiansen}, {Cecconi}, {Buchhave}, {Boschin}, \& {Andreuzzi}}]{Cloutier2021}
{Cloutier}, R., {Charbonneau}, D., {Stassun}, K.~G., {et~al.} 2021, \aj, 162, 79, \dodoi{10.3847/1538-3881/ac0157}

\bibitem[{Collaboration {et~al.}(2025)Collaboration, Pallathadka, Aghakhanloo, Aird, Almeida, Amrita, Anders, Anderson, Arseneau, Avila, Aviram, Aydar, Badenes, Barrera-Ballesteros, Bauer, Behmard, Berg, Besser, Bidin, Bizyaev, Blanc, Blanton, Bovy, Brandt, Brownstein, Buchner, Bulbul, Burchett, Carigi, Carlberg, Casey, Chakraborty, Chanamé, Chandra, Chiappini, Chilingarian, Comparat, Covey, Crumpler, Cunha, D'Onghia, Dai, Darling, Davis, Lee, Deacon, Delgado, Demasi, Demianenko, Demke, Donor, Drory, Durango, Dwelly, Egorov, Egorova, El-Badry, Eracleous, Fan, Farr, Finkbeiner, Fries, Frinchaboy, Fusillo, Félix, Gaensicke, Galligan, García, Gelfand, Grabowski, Grebel, Green, Greve, Grier, Griffith, Guetzoyan, Gupta, Hackshaw, Hall, Hawkins, Hegedűs, Hekker, Herbst, Hermes, Hernández-García, Hiremath, Hogg, Holtzman, Horne, Horta, Huang, Hutchinson, Häberle, Ibarra-Medel, Ji, Jofre, Johnson, Johnson, Johnston, Kaldor, Katkov, Khalatyan, Khoperskov, Klessen, Kluge, Koekemoer, Kollmeier, Kounkel, Kreckel,
  Krishnarao, Krumpe, Lacerna, Laporte, Lepine, Li, Liang, Limberg, Liu, Loebman, Long, Lu, Lucey, Lugo-Aranda, Martinez-Aldama, McKinnon, Medan, Merloni, Morrison, Myers, Mészáros, Müller-Horn, Nepal, Ness, Nidever, Nitschelm, Oravetz, Otto, Pan, Paolino, Peñaloza, Pinsonneault, Popp, Price-Whelan, Pulatova, Queiroz, Raddick, Rankine, Rix, Román-Zúñiga, Rosso, Runnoe, Saad, Salvato, Sanchez, Sattler, Saydjari, Sayres, Schlaufman, Schneider, Schwope, Seaton, Seeburger, Serna, Sharma, Shen, Sinha, Sizemore, Sniegowska, Song, Souto, Stassun, Steinmetz, Stone, Stone-Martinez, Stringfellow, Sánchez, Sánchez-Gallego, Tan, Tayar, Thai, Thakar, Thibodeaux, Ting, Tkachenko, Trakhtenbrot, Trincado, Troup, Trump, Ulloa, der Marel, Vera, Villanova, Villaseñor, Wang, Way, Weijmans, Wheeler, Wilson, Wofford, Wong, Wu, Wylezalek, Xue, Yan, Yang, Zakamska, Zari, Zasowski, Zeltyn, Zheng, Zucker, \& de~J.~Zermeño}]{sdsscollaboration2025}
Collaboration, S., Pallathadka, G.~A., Aghakhanloo, M., {et~al.} 2025, The Nineteenth Data Release of the Sloan Digital Sky Survey.
\newblock \doarXiv{2507.07093}

\bibitem[{{Collins} {et~al.}(2017){Collins}, {Kielkopf}, {Stassun}, \& {Hessman}}]{Collins:2017}
{Collins}, K.~A., {Kielkopf}, J.~F., {Stassun}, K.~G., \& {Hessman}, F.~V. 2017, \aj, 153, 77, \dodoi{10.3847/1538-3881/153/2/77}

\bibitem[{{Dekany} {et~al.}(2013){Dekany}, {Roberts}, {Burruss}, {Bouchez}, {Truong}, {Baranec}, {Guiwits}, {Hale}, {Angione}, {Trinh}, {Zolkower}, {Shelton}, {Palmer}, {Henning}, {Croner}, {Troy}, {McKenna}, {Tesch}, {Hildebrandt}, \& {Milburn}}]{dekany2013}
{Dekany}, R., {Roberts}, J., {Burruss}, R., {et~al.} 2013, \apj, 776, 130, \dodoi{10.1088/0004-637X/776/2/130}

\bibitem[{{Demangeon} {et~al.}(2021){Demangeon}, {Zapatero Osorio}, {Alibert}, {Barros}, {Adibekyan}, {Tabernero}, {Antoniadis-Karnavas}, {Camacho}, {Su{\'a}rez Mascare{\~n}o}, {Oshagh}, {Micela}, {Sousa}, {Lovis}, {Pepe}, {Rebolo}, {Cristiani}, {Santos}, {Allart}, {Allende Prieto}, {Bossini}, {Bouchy}, {Cabral}, {Damasso}, {Di Marcantonio}, {D'Odorico}, {Ehrenreich}, {Faria}, {Figueira}, {G{\'e}nova Santos}, {Haldemann}, {Hara}, {Gonz{\'a}lez Hern{\'a}ndez}, {Lavie}, {Lillo-Box}, {Lo Curto}, {Martins}, {M{\'e}gevand}, {Mehner}, {Molaro}, {Nunes}, {Pall{\'e}}, {Pasquini}, {Poretti}, {Sozzetti}, \& {Udry}}]{Demangeon2021}
{Demangeon}, O.~D.~S., {Zapatero Osorio}, M.~R., {Alibert}, Y., {et~al.} 2021, \aap, 653, A41, \dodoi{10.1051/0004-6361/202140728}

\bibitem[{{Demory} {et~al.}(2020){Demory}, {Pozuelos}, {G{\'o}mez Maqueo Chew}, {Sabin}, {Petrucci}, {Schroffenegger}, {Grimm}, {Sestovic}, {Gillon}, {McCormac}, {Barkaoui}, {Benz}, {Bieryla}, {Bouchy}, {Burdanov}, {Collins}, {de Wit}, {Dressing}, {Garcia}, {Giacalone}, {Guerra}, {Haldemann}, {Heng}, {Jehin}, {Jofr{\'e}}, {Kane}, {Lillo-Box}, {Maign{\'e}}, {Mordasini}, {Morris}, {Niraula}, {Queloz}, {Rackham}, {Savel}, {Soubkiou}, {Srdoc}, {Stassun}, {Triaud}, {Zambelli}, {Ricker}, {Latham}, {Seager}, {Winn}, {Jenkins}, {Calvario-Vel{\'a}squez}, {Franco Herrera}, {Colorado}, {Cadena Zepeda}, {Figueroa}, {Watson}, {Lugo-Ibarra}, {Carigi}, {Guisa}, {Herrera}, {Sierra D{\'\i}az}, {Su{\'a}rez}, {Barrado}, {Batalha}, {Benkhaldoun}, {Chontos}, {Dai}, {Essack}, {Ghachoui}, {Huang}, {Huber}, {Isaacson}, {Lissauer}, {Morales-Calder{\'o}n}, {Robertson}, {Roy}, {Twicken}, {Vanderburg}, \& {Weiss}}]{Demory2020}
{Demory}, B.~O., {Pozuelos}, F.~J., {G{\'o}mez Maqueo Chew}, Y., {et~al.} 2020, \aap, 642, A49, \dodoi{10.1051/0004-6361/202038616}

\bibitem[{{Fleming} {et~al.}(2020){Fleming}, {Barnes}, {Luger}, \& {VanderPlas}}]{Fleming2020}
{Fleming}, D.~P., {Barnes}, R., {Luger}, R., \& {VanderPlas}, J.~T. 2020, \apj, 891, 155, \dodoi{10.3847/1538-4357/ab77ad}

\bibitem[{{Fortune} {et~al.}(2025){Fortune}, {Gibson}, {Diamond-Lowe}, {Mendon{\c{c}}a}, {Gressier}, {Kitzmann}, {Allen}, {August}, {Ih}, {Meier Vald{\'e}s}, {Zgraggen}, {Buchhave}, {Demory}, {Espinoza}, {Heng}, {Jones}, \& {Rathcke}}]{Fortune2025}
{Fortune}, M., {Gibson}, N.~P., {Diamond-Lowe}, H., {et~al.} 2025, arXiv e-prints, arXiv:2505.22186, \dodoi{10.48550/arXiv.2505.22186}

\bibitem[{{Furlan} {et~al.}(2017){Furlan}, {Ciardi}, {Everett}, {Saylors}, {Teske}, {Horch}, {Howell}, {van Belle}, {Hirsch}, {Gautier}, {Adams}, {Barrado}, {Cartier}, {Dressing}, {Dupree}, {Gilliland}, {Lillo-Box}, {Lucas}, \& {Wang}}]{furlan2017}
{Furlan}, E., {Ciardi}, D.~R., {Everett}, M.~E., {et~al.} 2017, \aj, 153, 71, \dodoi{10.3847/1538-3881/153/2/71}

\bibitem[{{Gan} {et~al.}(2020){Gan}, {Shporer}, {Livingston}, {Collins}, {Mao}, {Trani}, {Gandolfi}, {Hirano}, {Luque}, {Stassun}, {Ziegler}, {Howell}, {Hellier}, {Irwin}, {Winters}, {Anderson}, {Brice{\~n}o}, {Law}, {Mann}, {Bonfils}, {Astudillo-Defru}, {Jensen}, {Anglada-Escud{\'e}}, {Ricker}, {Vanderspek}, {Latham}, {Seager}, {Winn}, {Jenkins}, {Furesz}, {Guerrero}, {Quintana}, {Twicken}, {Caldwell}, {Tenenbaum}, {Huang}, {Rowden}, \& {Rojas-Ayala}}]{Gan2020}
{Gan}, T., {Shporer}, A., {Livingston}, J.~H., {et~al.} 2020, \aj, 159, 160, \dodoi{10.3847/1538-3881/ab775a}

\bibitem[{{Giacalone} {et~al.}(2021){Giacalone}, {Dressing}, {Jensen}, {Collins}, {Ricker}, {Vanderspek}, {Seager}, {Winn}, {Jenkins}, {Barclay}, {Barkaoui}, {Cadieux}, {Charbonneau}, {Collins}, {Conti}, {Doyon}, {Evans}, {Ghachoui}, {Gillon}, {Guerrero}, {Hart}, {Jehin}, {Kielkopf}, {McLean}, {Murgas}, {Palle}, {Parviainen}, {Pozuelos}, {Relles}, {Shporer}, {Socia}, {Stockdale}, {Tan}, {Torres}, {Twicken}, {Waalkes}, \& {Waite}}]{Giacalone2021}
{Giacalone}, S., {Dressing}, C.~D., {Jensen}, E. L.~N., {et~al.} 2021, \aj, 161, 24, \dodoi{10.3847/1538-3881/abc6af}

\bibitem[{{Gomez Barrientos} {et~al.}(2025){Gomez Barrientos}, {Greklek-McKeon}, {Knutson}, {Giacalone}, {Levine}, {Saidel}, {Vissapragada}, {Ciardi}, {Collins}, {Latham}, {Watkins}, {Budnikova}, {Cheryasov}, {Fukui}, {Bieryla}, {Shporer}, {Tofflemire}, {Clark}, {Stockdale}, {Littlefield}, {Gilbert}, {Palle}, {Girardin}, {Murgas}, {Bergsten}, {Osborn}, {Crossfield}, {de Leon}, {Higuera}, {Isogai}, {Everett}, {Lund}, {Narita}, {Schwarz}, {Zambelli}, \& {Howell}}]{GomezBarrientos2025}
{Gomez Barrientos}, J., {Greklek-McKeon}, M., {Knutson}, H.~A., {et~al.} 2025, \aj, 170, 148, \dodoi{10.3847/1538-3881/ade68b}

\bibitem[{{G{\'o}mez Maqueo Chew} {et~al.}(2023){G{\'o}mez Maqueo Chew}, {Demory}, {Sabin}, {Wells}, {Petrucci}, {Schroffenegger}, {G{\'o}mez-Mu{\~n}oz}, {Schanche}, \& {Saint-Ex Team}}]{yilen2023}
{G{\'o}mez Maqueo Chew}, Y., {Demory}, B.~O., {Sabin}, L., {et~al.} 2023, in Revista Mexicana de Astronomia y Astrofisica Conference Series, Vol.~55, Revista Mexicana de Astronomia y Astrofisica Conference Series, 44--46, \dodoi{10.22201/ia.14052059p.2023.55.10}

\bibitem[{{Grant} \& {Wakeford}(2024)}]{Grant2024}
{Grant}, D., \& {Wakeford}, H. 2024, The Journal of Open Source Software, 9, 6816, \dodoi{10.21105/joss.06816}

\bibitem[{{Greene} {et~al.}(2023){Greene}, {Bell}, {Ducrot}, {Dyrek}, {Lagage}, \& {Fortney}}]{Greene2023}
{Greene}, T.~P., {Bell}, T.~J., {Ducrot}, E., {et~al.} 2023, \nat, 618, 39, \dodoi{10.1038/s41586-023-05951-7}

\bibitem[{{Greklek-McKeon} {et~al.}(2023){Greklek-McKeon}, {Knutson}, {Vissapragada}, {Jontof-Hutter}, {Chachan}, {Thorngren}, \& {Vasisht}}]{Greklek-McKeon2023}
{Greklek-McKeon}, M., {Knutson}, H.~A., {Vissapragada}, S., {et~al.} 2023, \aj, 165, 48, \dodoi{10.3847/1538-3881/ac8553}

\bibitem[{{Guerrero} {et~al.}(2021){Guerrero}, {Seager}, {Huang}, {Vanderburg}, {Garcia Soto}, {Mireles}, {Hesse}, {Fong}, {Glidden}, {Shporer}, {Latham}, {Collins}, {Quinn}, {Burt}, {Dragomir}, {Crossfield}, {Vanderspek}, {Fausnaugh}, {Burke}, {Ricker}, {Daylan}, {Essack}, {G{\"u}nther}, {Osborn}, {Pepper}, {Rowden}, {Sha}, {Villanueva}, {Yahalomi}, {Yu}, {Ballard}, {Batalha}, {Berardo}, {Chontos}, {Dittmann}, {Esquerdo}, {Mikal-Evans}, {Jayaraman}, {Krishnamurthy}, {Louie}, {Mehrle}, {Niraula}, {Rackham}, {Rodriguez}, {Rowden}, {Sousa-Silva}, {Watanabe}, {Wong}, {Zhan}, {Zivanovic}, {Christiansen}, {Ciardi}, {Swain}, {Lund}, {Mullally}, {Fleming}, {Rodriguez}, {Boyd}, {Quintana}, {Barclay}, {Col{\'o}n}, {Rinehart}, {Schlieder}, {Clampin}, {Jenkins}, {Twicken}, {Caldwell}, {Coughlin}, {Henze}, {Lissauer}, {Morris}, {Rose}, {Smith}, {Tenenbaum}, {Ting}, {Wohler}, {Bakos}, {Bean}, {Berta-Thompson}, {Bieryla}, {Bouma}, {Buchhave}, {Butler}, {Charbonneau}, {Doty}, {Ge}, {Holman}, {Howard}, {Kaltenegger}, {Kane},
  {Kjeldsen}, {Kreidberg}, {Lin}, {Minsky}, {Narita}, {Paegert}, {P{\'a}l}, {Palle}, {Sasselov}, {Spencer}, {Sozzetti}, {Stassun}, {Torres}, {Udry}, \& {Winn}}]{Guerrero2021}
{Guerrero}, N.~M., {Seager}, S., {Huang}, C.~X., {et~al.} 2021, \apjs, 254, 39, \dodoi{10.3847/1538-4365/abefe1}

\bibitem[{{Hallatt} \& {Lee}(2025)}]{Hallatt2025}
{Hallatt}, T., \& {Lee}, E.~J. 2025, \apj, 979, 120, \dodoi{10.3847/1538-4357/ad9aa1}

\bibitem[{{Hayward} {et~al.}(2001){Hayward}, {Brandl}, {Pirger}, {Blacken}, {Gull}, {Schoenwald}, \& {Houck}}]{hayward2001}
{Hayward}, T.~L., {Brandl}, B., {Pirger}, B., {et~al.} 2001, \pasp, 113, 105, \dodoi{10.1086/317969}

\bibitem[{{Henry} {et~al.}(1994){Henry}, {Kirkpatrick}, \& {Simons}}]{Henry1994}
{Henry}, T.~J., {Kirkpatrick}, J.~D., \& {Simons}, D.~A. 1994, \aj, 108, 1437, \dodoi{10.1086/117167}

\bibitem[{{Hoffman} \& {Gelman}(2011)}]{Hoffman2011}
{Hoffman}, M.~D., \& {Gelman}, A. 2011, arXiv e-prints, arXiv:1111.4246, \dodoi{10.48550/arXiv.1111.4246}

\bibitem[{{Holczer} {et~al.}(2016){Holczer}, {Mazeh}, {Nachmani}, {Jontof-Hutter}, {Ford}, {Fabrycky}, {Ragozzine}, {Kane}, \& {Steffen}}]{Holczer2016}
{Holczer}, T., {Mazeh}, T., {Nachmani}, G., {et~al.} 2016, \apjs, 225, 9, \dodoi{10.3847/0067-0049/225/1/9}

\bibitem[{{Hori} \& {Ogihara}(2020)}]{Hori2020}
{Hori}, Y., \& {Ogihara}, M. 2020, \apj, 889, 77, \dodoi{10.3847/1538-4357/ab6168}

\bibitem[{{Huang} {et~al.}(2020{\natexlab{a}}){Huang}, {Vanderburg}, {P{\'a}l}, {Sha}, {Yu}, {Fong}, {Fausnaugh}, {Shporer}, {Guerrero}, {Vanderspek}, \& {Ricker}}]{Huang2020}
{Huang}, C.~X., {Vanderburg}, A., {P{\'a}l}, A., {et~al.} 2020{\natexlab{a}}, Research Notes of the American Astronomical Society, 4, 204, \dodoi{10.3847/2515-5172/abca2e}

\bibitem[{{Huang} {et~al.}(2020{\natexlab{b}}){Huang}, {Vanderburg}, {P{\'a}l}, {Sha}, {Yu}, {Fong}, {Fausnaugh}, {Shporer}, {Guerrero}, {Vanderspek}, \& {Ricker}}]{Huang2020b}
---. 2020{\natexlab{b}}, Research Notes of the American Astronomical Society, 4, 206, \dodoi{10.3847/2515-5172/abca2d}

\bibitem[{{Huber} {et~al.}(2017){Huber}, {Zinn}, {Bojsen-Hansen}, {Pinsonneault}, {Sahlholdt}, {Serenelli}, {Silva Aguirre}, {Stassun}, {Stello}, {Tayar}, {Bastien}, {Bedding}, {Buchhave}, {Chaplin}, {Davies}, {Garc{\'\i}a}, {Latham}, {Mathur}, {Mosser}, \& {Sharma}}]{Huber2017}
{Huber}, D., {Zinn}, J., {Bojsen-Hansen}, M., {et~al.} 2017, \apj, 844, 102, \dodoi{10.3847/1538-4357/aa75ca}

\bibitem[{{Jenkins}(2002)}]{Jenkins2002}
{Jenkins}, J.~M. 2002, \apj, 575, 493, \dodoi{10.1086/341136}

\bibitem[{{Jenkins} {et~al.}(2020){Jenkins}, {Tenenbaum}, {Seader}, {Burke}, {McCauliff}, {Smith}, {Twicken}, \& {Chandrasekaran}}]{Jenkins2020}
{Jenkins}, J.~M., {Tenenbaum}, P., {Seader}, S., {et~al.} 2020, {Kepler Data Processing Handbook: Transiting Planet Search}, Kepler Science Document KSCI-19081-003

\bibitem[{{Jenkins} {et~al.}(2010){Jenkins}, {Chandrasekaran}, {McCauliff}, {Caldwell}, {Tenenbaum}, {Li}, {Klaus}, {Cote}, \& {Middour}}]{Jenkins2010}
{Jenkins}, J.~M., {Chandrasekaran}, H., {McCauliff}, S.~D., {et~al.} 2010, in Society of Photo-Optical Instrumentation Engineers (SPIE) Conference Series, Vol. 7740, Software and Cyberinfrastructure for Astronomy, ed. N.~M. {Radziwill} \& A.~{Bridger}, 77400D, \dodoi{10.1117/12.856764}

\bibitem[{{Jenkins} {et~al.}(2016){Jenkins}, {Twicken}, {McCauliff}, {Campbell}, {Sanderfer}, {Lung}, {Mansouri-Samani}, {Girouard}, {Tenenbaum}, {Klaus}, {Smith}, {Caldwell}, {Chacon}, {Henze}, {Heiges}, {Latham}, {Morgan}, {Swade}, {Rinehart}, \& {Vanderspek}}]{Jenkins2016}
{Jenkins}, J.~M., {Twicken}, J.~D., {McCauliff}, S., {et~al.} 2016, in Society of Photo-Optical Instrumentation Engineers (SPIE) Conference Series, Vol. 9913, Software and Cyberinfrastructure for Astronomy IV, ed. G.~{Chiozzi} \& J.~C. {Guzman}, 99133E, \dodoi{10.1117/12.2233418}

\bibitem[{{Katz} {et~al.}(2023){Katz}, {Sartoretti}, {Guerrier}, {Panuzzo}, {Seabroke}, {Th{\'e}venin}, {Cropper}, {Benson}, {Blomme}, {Haigron}, {Marchal}, {Smith}, {Baker}, {Chemin}, {Damerdji}, {David}, {Dolding}, {Fr{\'e}mat}, {Gosset}, {Jan{\ss}en}, {Jasniewicz}, {Lobel}, {Plum}, {Samaras}, {Snaith}, {Soubiran}, {Vanel}, {Zwitter}, {Antoja}, {Arenou}, {Babusiaux}, {Brouillet}, {Caffau}, {Di Matteo}, {Fabre}, {Fabricius}, {Fragkoudi}, {Haywood}, {Huckle}, {Hottier}, {Lasne}, {Leclerc}, {Mastrobuono-Battisti}, {Royer}, {Teyssier}, {Zorec}, {Crifo}, {Jean-Antoine Piccolo}, {Turon}, \& {Viala}}]{Katz2023}
{Katz}, D., {Sartoretti}, P., {Guerrier}, A., {et~al.} 2023, \aap, 674, A5, \dodoi{10.1051/0004-6361/202244220}

\bibitem[{{Kempton} {et~al.}(2018){Kempton}, {Bean}, {Louie}, {Deming}, {Koll}, {Mansfield}, {Christiansen}, {L{\'o}pez-Morales}, {Swain}, {Zellem}, {Ballard}, {Barclay}, {Barstow}, {Batalha}, {Beatty}, {Berta-Thompson}, {Birkby}, {Buchhave}, {Charbonneau}, {Cowan}, {Crossfield}, {de Val-Borro}, {Doyon}, {Dragomir}, {Gaidos}, {Heng}, {Hu}, {Kane}, {Kreidberg}, {Mallonn}, {Morley}, {Narita}, {Nascimbeni}, {Pall{\'e}}, {Quintana}, {Rauscher}, {Seager}, {Shkolnik}, {Sing}, {Sozzetti}, {Stassun}, {Valenti}, \& {von Essen}}]{Kempton2018}
{Kempton}, E. M.~R., {Bean}, J.~L., {Louie}, D.~R., {et~al.} 2018, \pasp, 130, 114401, \dodoi{10.1088/1538-3873/aadf6f}

\bibitem[{{Kipping}(2013)}]{Kipping2013}
{Kipping}, D.~M. 2013, \mnras, 434, L51, \dodoi{10.1093/mnrasl/slt075}

\bibitem[{{Kirkpatrick} {et~al.}(1999){Kirkpatrick}, {Reid}, {Liebert}, {Cutri}, {Nelson}, {Beichman}, {Dahn}, {Monet}, {Gizis}, \& {Skrutskie}}]{Kirkpatrick1999}
{Kirkpatrick}, J.~D., {Reid}, I.~N., {Liebert}, J., {et~al.} 1999, \apj, 519, 802, \dodoi{10.1086/307414}

\bibitem[{Kollmeier {et~al.}(2025)Kollmeier, Rix, Aerts, Aird, Alfaro, Almeida, Anderson, Óscar Jiménez~Arranz, Arseneau, Assef, Aviram, Aydar, Badenes, Bandyopadhyay, Barger, Barkhouser, Bauer, Bender, Besser, Bhattarai, Bilgi, Bird, Bizyaev, Blanc, Blanton, Bochanski, Bovy, Brandon, Brandt, Brownstein, Buchner, Burchett, Carlberg, Casey, Castaneda-Carlos, Chakraborty, Chanamé, Chandra, Cherinka, Chilingarian, Comparat, Cosens, Covey, Crane, Crumpler, Cunha, Cunningham, Dai, Darling, Jr., Davis, Lee, Deacon, Delgado, Demasi, Demianenko, Derwent, D'Onghia, Mille, Dias, Donor, Drory, Dwelly, Egorov, Egorova, El-Badry, Engelman, Eracleous, Fan, Farr, Fries, Frinchaboy, Froning, Gänsicke, García, Gelfand, Fusillo, Glover, Grabowski, Grebel, Green, Grier, Gupta, Gray, Häberle, Hall, Hammond, Hawkins, Harding, Hegedűs, Herbst, Hermes, Hidalgo, Hilder, Hogg, Holtzman, Horta, Huang, Hwang, Ibarra-Medel, Imig, Inight, Jana, Ji, Jofre, Johns, Johnson, Johnson, Johnston, Jones, Katkov, Koekemoer, Kounkel,
  Kreckel, Krishnarao, Krumpe, Kumari, Kupfer, Lacerna, Laporte, Lepine, Li, Liu, Loebman, Long, Roman-Lopes, Lu, Majewski, Maoz, McKinnon, Medan, Merloni, Minniti, Morrison, Myers, Mészáros, Nandra, Nayak, Ness, Nidever, O'Brien, Oeur, Oravetz, Oravetz, Otto, Pallathadka, Palunas, Pan, Pappalardo, Pandey, Peñaloza, Pinsonneault, Pogge, Popp, Price-Whelan, Pulatova, Qiu, Ramirez, Rankine, Ricci, Runnoe, Sanchez, Salvato, Sattler, Saydjari, Sayres, Schlaufman, Schneider, Schreiber, Schwope, Serna, Shen, Sifón, Singh, Sinha, Smee, Song, Souto, Stassun, Steinmetz, Stone-Martinez, Stringfellow, Stutz, José, Sá, nchez Gallego, Tan, Tayar, Thai, Thakar, Ting, Tkachenko, Tovmasian, Trakhtenbrot, Fernández-Trincado, Troup, Trump, Tuttle, van~der Marel, Villanova, Villaseñor, Wachter, Way, Weijmans, Weinberg, Wheeler, Wilson, Wiggins, Wong, Wu, Wylezalek, Xue, Yang, Zakamska, Zari, Zasowski, Zeltyn, Zucker, Zúñiga, \& de~J.~Zermeño}]{sloandigitalskysurveyv2025}
Kollmeier, J.~A., Rix, H.-W., Aerts, C., {et~al.} 2025, Sloan Digital Sky Survey-V: Pioneering Panoptic Spectroscopy.
\newblock \doarXiv{2507.06989}

\bibitem[{{Kov{\'a}cs} {et~al.}(2002){Kov{\'a}cs}, {Zucker}, \& {Mazeh}}]{Kovacs2002}
{Kov{\'a}cs}, G., {Zucker}, S., \& {Mazeh}, T. 2002, \aap, 391, 369, \dodoi{10.1051/0004-6361:20020802}

\bibitem[{{Kreidberg} \& {Stevenson}(2025)}]{Kreidberg2025}
{Kreidberg}, L., \& {Stevenson}, K.~B. 2025, arXiv e-prints, arXiv:2507.00933, \dodoi{10.48550/arXiv.2507.00933}

\bibitem[{{Kunimoto} \& {Daylan}(2021)}]{Kunimoto2021}
{Kunimoto}, M., \& {Daylan}, T. 2021, in Posters from the TESS Science Conference II (TSC2), 62, \dodoi{10.5281/zenodo.5125527}

\bibitem[{{Lightkurve Collaboration} {et~al.}(2018){Lightkurve Collaboration}, {Cardoso}, {Hedges}, {Gully-Santiago}, {Saunders}, {Cody}, {Barclay}, {Hall}, {Sagear}, {Turtelboom}, {Zhang}, {Tzanidakis}, {Mighell}, {Coughlin}, {Bell}, {Berta-Thompson}, {Williams}, {Dotson}, \& {Barentsen}}]{2018ascl.soft12013L}
{Lightkurve Collaboration}, {Cardoso}, J.~V.~d.~M., {Hedges}, C., {et~al.} 2018, {Lightkurve: Kepler and TESS time series analysis in Python}, Astrophysics Source Code Library.
\newblock \doeprint{1812.013}

\bibitem[{{Lomb}(1976)}]{Lomb1976}
{Lomb}, N.~R. 1976, \apss, 39, 447, \dodoi{10.1007/BF00648343}

\bibitem[{{Luger} {et~al.}(2019){Luger}, {Agol}, {Foreman-Mackey}, {Fleming}, {Lustig-Yaeger}, \& {Deitrick}}]{Luger2019}
{Luger}, R., {Agol}, E., {Foreman-Mackey}, D., {et~al.} 2019, \aj, 157, 64, \dodoi{10.3847/1538-3881/aae8e5}

\bibitem[{{Luque} \& {Pall{\'e}}(2022)}]{Luque2022}
{Luque}, R., \& {Pall{\'e}}, E. 2022, Science, 377, 1211, \dodoi{10.1126/science.abl7164}

\bibitem[{{Luque} {et~al.}(2025){Luque}, {Coy}, {Xue}, {Feinstein}, {Ahrer}, {Changeat}, {Zhang}, {Moran}, {Bean}, {Kite}, {Weiner Mansfield}, \& {Pall{\'e}}}]{Luque2025}
{Luque}, R., {Coy}, B.~P., {Xue}, Q., {et~al.} 2025, \aj, 170, 49, \dodoi{10.3847/1538-3881/addb40}

\bibitem[{{Mann} {et~al.}(2015){Mann}, {Feiden}, {Gaidos}, {Boyajian}, \& {von Braun}}]{Mann2015}
{Mann}, A.~W., {Feiden}, G.~A., {Gaidos}, E., {Boyajian}, T., \& {von Braun}, K. 2015, \apj, 804, 64, \dodoi{10.1088/0004-637X/804/1/64}

\bibitem[{{Mann} {et~al.}(2019){Mann}, {Dupuy}, {Kraus}, {Gaidos}, {Ansdell}, {Ireland}, {Rizzuto}, {Hung}, {Dittmann}, {Factor}, {Feiden}, {Martinez}, {Ru{\'\i}z-Rodr{\'\i}guez}, \& {Thao}}]{Mann2019}
{Mann}, A.~W., {Dupuy}, T., {Kraus}, A.~L., {et~al.} 2019, \apj, 871, 63, \dodoi{10.3847/1538-4357/aaf3bc}

\bibitem[{{McCully} {et~al.}(2018){McCully}, {Volgenau}, {Harbeck}, {Lister}, {Saunders}, {Turner}, {Siiverd}, \& {Bowman}}]{McCully:2018}
{McCully}, C., {Volgenau}, N.~H., {Harbeck}, D.-R., {et~al.} 2018, in Society of Photo-Optical Instrumentation Engineers (SPIE) Conference Series, Vol. 10707, \procspie, 107070K, \dodoi{10.1117/12.2314340}

\bibitem[{{Meier Vald{\'e}s} {et~al.}(2025){Meier Vald{\'e}s}, {Demory}, {Diamond-Lowe}, {Mendon{\c{c}}a}, {August}, {Fortune}, {Allen}, {Kitzmann}, {Gressier}, {Hooton}, {Jones}, {Buchhave}, {Espinoza}, {Fisher}, {Gibson}, {Heng}, {Hoeijmakers}, {Prinoth}, {Rathcke}, \& {Eastman}}]{MeierValdes2025}
{Meier Vald{\'e}s}, E.~A., {Demory}, B.~O., {Diamond-Lowe}, H., {et~al.} 2025, \aap, 698, A68, \dodoi{10.1051/0004-6361/202453449}

\bibitem[{Melo {et~al.}(2024)Melo, Souto, Cunha, Smith, Wanderley, Grilo, Camara, Murta, Hejazi, Crossfield, Teske, Luque, Zhang, \& Bean}]{Melo2024}
Melo, E., Souto, D., Cunha, K., {et~al.} 2024, Monthly Notices of the Royal Astronomical Society, 2406, 00111, \dodoi{10.48550/arXiv.2406.00111}

\bibitem[{{Narita} {et~al.}(2020){Narita}, {Fukui}, {Yamamuro}, {Harbeck}, {Bowman}, {Elphick}, {Nation}, {Armstrong}, {Han}, {Abe}, {Ikoma}, {Isogai}, {Kawauchi}, {Kurita}, {Kusakabe}, {de Leon}, {Livingston}, {Mori}, {Nishiumi}, {Tamura}, {Watanabe}, {Volgenau}, {Heinrich-Josties}, {Foale}, {Daily}, {McCully}, {Kirby}, {Smith}, {Haworth}, {Conway}, {Storrie-Lombardi}, {Rosing}, {Chatelain}, {Bachelet}, {Johnson}, \& {Rabus}}]{Narita:2020}
{Narita}, N., {Fukui}, A., {Yamamuro}, T., {et~al.} 2020, in Society of Photo-Optical Instrumentation Engineers (SPIE) Conference Series, Vol. 11447, Society of Photo-Optical Instrumentation Engineers (SPIE) Conference Series, 114475K, \dodoi{10.1117/12.2559947}

\bibitem[{{NASA Exoplanet Archive}(2024)}]{ps}
{NASA Exoplanet Archive}. 2024, Planetary Systems, Version: YYYY-MM-DD HH:MM,  NExScI-Caltech/IPAC, \dodoi{10.26133/NEA12}

\bibitem[{{Ness} {et~al.}(2015){Ness}, {Hogg}, {Rix}, {Ho}, \& {Zasowski}}]{Ness2015}
{Ness}, M., {Hogg}, D.~W., {Rix}, H.~W., {Ho}, A. Y.~Q., \& {Zasowski}, G. 2015, \apj, 808, 16, \dodoi{10.1088/0004-637X/808/1/16}

\bibitem[{Nidever {et~al.}(2015)Nidever, Holtzman, Prieto, Beland, Bender, Bizyaev, Burton, Desphande, Fleming, Pérez, Hearty, Majewski, Mészáros, Muna, Nguyen, Schiavon, Shetrone, Skrutskie, Sobeck, \& Wilson}]{Nidever_2015}
Nidever, D.~L., Holtzman, J.~A., Prieto, C.~A., {et~al.} 2015, The Astronomical Journal, 150, 173, \dodoi{10.1088/0004-6256/150/6/173}

\bibitem[{{Nutzman} \& {Charbonneau}(2008)}]{Nutzman2008}
{Nutzman}, P., \& {Charbonneau}, D. 2008, \pasp, 120, 317, \dodoi{10.1086/533420}

\bibitem[{{Pass} {et~al.}(2025){Pass}, {Charbonneau}, \& {Vanderburg}}]{Pass2025}
{Pass}, E.~K., {Charbonneau}, D., \& {Vanderburg}, A. 2025, \apjl, 986, L3, \dodoi{10.3847/2041-8213/adda39}

\bibitem[{{Roettenbacher} \& {Kane}(2017)}]{Roettenbacher2017}
{Roettenbacher}, R.~M., \& {Kane}, S.~R. 2017, \apj, 851, 77, \dodoi{10.3847/1538-4357/aa991e}

\bibitem[{{Ross} {et~al.}(2025){Ross}, {Reggiani}, {Schlaufman}, {Plotnykov}, \& {Valencia}}]{Ross2025}
{Ross}, A., {Reggiani}, H., {Schlaufman}, K.~C., {Plotnykov}, M., \& {Valencia}, D. 2025, arXiv e-prints, arXiv:2508.16421, \dodoi{10.48550/arXiv.2508.16421}

\bibitem[{Rowe {et~al.}(2014)Rowe, Bryson, Marcy, Lissauer, Jontof-Hutter, Mullally, Gilliland, Issacson, Ford, Howell, Borucki, Haas, Huber, Steffen, Thompson, Quintana, Barclay, Still, Fortney, Gautier, Hunter, Caldwell, Ciardi, Devore, Cochran, Jenkins, Agol, Carter, \& Geary}]{Rowe2014}
Rowe, J.~F., Bryson, S.~T., Marcy, G.~W., {et~al.} 2014, The Astrophysical Journal, 784, 45, \dodoi{10.1088/0004-637X/784/1/45}

\bibitem[{{Scargle}(1982)}]{Scargle1982}
{Scargle}, J.~D. 1982, \apj, 263, 835, \dodoi{10.1086/160554}

\bibitem[{Schwarz(1978)}]{Schwarz1978}
Schwarz, G. 1978, The Annals of Statistics, 6, 461 , \dodoi{10.1214/aos/1176344136}

\bibitem[{{Scott} {et~al.}(2025){Scott}, {Triaud}, {Barkaoui}, {Sebastian}, {Burgasser}, {Collins}, {Dransfield}, {Hellier}, {Howell}, {Piette}, {Rackham}, {Stassun}, {Stokholm}, {Timmermans}, {Watkins}, {Fausnaugh}, {Fukui}, {Jenkins}, {Narita}, {Ricker}, {Softich}, {Schwarz}, {Seager}, {Shporer}, {Theissen}, {Twicken}, {Winn}, \& {Watanabe}}]{Scott2025}
{Scott}, M.~G., {Triaud}, A. H.~M.~J., {Barkaoui}, K., {et~al.} 2025, \mnras, 540, 1909, \dodoi{10.1093/mnras/staf684}

\bibitem[{{See} {et~al.}(2025){See}, {Fairman}, {Amard}, \& {Hall}}]{See2025}
{See}, V., {Fairman}, C., {Amard}, L., \& {Hall}, O. 2025, \mnras, \dodoi{10.1093/mnras/staf1753}

\bibitem[{{See} {et~al.}(2024){See}, {Lu}, {Amard}, \& {Roquette}}]{See2024}
{See}, V., {Lu}, Y.~L., {Amard}, L., \& {Roquette}, J. 2024, \mnras, 533, 1290, \dodoi{10.1093/mnras/stae1828}

\bibitem[{{Smith} {et~al.}(2012){Smith}, {Stumpe}, {Van Cleve}, {Jenkins}, {Barclay}, {Fanelli}, {Girouard}, {Kolodziejczak}, {McCauliff}, {Morris}, \& {Twicken}}]{Smith2012}
{Smith}, J.~C., {Stumpe}, M.~C., {Van Cleve}, J.~E., {et~al.} 2012, \pasp, 124, 1000, \dodoi{10.1086/667697}

\bibitem[{Souto {et~al.}(2021)Souto, Cunha, \& Smith}]{Souto2021}
Souto, D., Cunha, K., \& Smith, V.~V. 2021, The Astrophysical Journal, 917, 11, \dodoi{10.3847/1538-4357/abfdb5}

\bibitem[{Souto {et~al.}(2020)Souto, Cunha, Smith, Allende~Prieto, Burgasser, Covey, García-Hernández, Holtzman, Johnson, Jönsson, Mahadevan, Majewski, Masseron, Shetrone, Rojas-Ayala, Sobeck, Stassun, Terrien, Teske, Wanderley, \& Zamora}]{Souto2020}
Souto, D., Cunha, K., Smith, V.~V., {et~al.} 2020, The Astrophysical Journal, 890, 133, \dodoi{10.3847/1538-4357/ab6d07}

\bibitem[{{Stassun} {et~al.}(2019){Stassun}, {Oelkers}, {Paegert}, {Torres}, {Pepper}, {De Lee}, {Collins}, {Latham}, {Muirhead}, {Chittidi}, {Rojas-Ayala}, {Fleming}, {Rose}, {Tenenbaum}, {Ting}, {Kane}, {Barclay}, {Bean}, {Brassuer}, {Charbonneau}, {Ge}, {Lissauer}, {Mann}, {McLean}, {Mullally}, {Narita}, {Plavchan}, {Ricker}, {Sasselov}, {Seager}, {Sharma}, {Shiao}, {Sozzetti}, {Stello}, {Vanderspek}, {Wallace}, \& {Winn}}]{Stassun2019}
{Stassun}, K.~G., {Oelkers}, R.~J., {Paegert}, M., {et~al.} 2019, \aj, 158, 138, \dodoi{10.3847/1538-3881/ab3467}

\bibitem[{{Stefansson} {et~al.}(2017){Stefansson}, {Mahadevan}, {Hebb}, {Wisniewski}, {Huehnerhoff}, {Morris}, {Halverson}, {Zhao}, {Wright}, {O'rourke}, {Knutson}, {Hawley}, {Kanodia}, {Li}, {Hagen}, {Liu}, {Beatty}, {Bender}, {Robertson}, {Dembicky}, {Gray}, {Ketzeback}, {McMillan}, \& {Rudyk}}]{Stefansson2017}
{Stefansson}, G., {Mahadevan}, S., {Hebb}, L., {et~al.} 2017, \apj, 848, 9, \dodoi{10.3847/1538-4357/aa88aa}

\bibitem[{{Stetson}(1987)}]{Stetson1987}
{Stetson}, P.~B. 1987, \pasp, 99, 191, \dodoi{10.1086/131977}

\bibitem[{{Strakhov} {et~al.}(2023){Strakhov}, {Safonov}, \& {Cheryasov}}]{Strakhov2023}
{Strakhov}, I.~A., {Safonov}, B.~S., \& {Cheryasov}, D.~V. 2023, Astrophysical Bulletin, 78, 234, \dodoi{10.1134/S1990341323020104}

\bibitem[{{Stumpe} {et~al.}(2014){Stumpe}, {Smith}, {Catanzarite}, {Van Cleve}, {Jenkins}, {Twicken}, \& {Girouard}}]{Stumpe2014}
{Stumpe}, M.~C., {Smith}, J.~C., {Catanzarite}, J.~H., {et~al.} 2014, \pasp, 126, 100, \dodoi{10.1086/674989}

\bibitem[{{Stumpe} {et~al.}(2012){Stumpe}, {Smith}, {Van Cleve}, {Twicken}, {Barclay}, {Fanelli}, {Girouard}, {Jenkins}, {Kolodziejczak}, {McCauliff}, \& {Morris}}]{Stumpe2012}
{Stumpe}, M.~C., {Smith}, J.~C., {Van Cleve}, J.~E., {et~al.} 2012, \pasp, 124, 985, \dodoi{10.1086/667698}

\bibitem[{{Vissapragada} {et~al.}(2020){Vissapragada}, {Jontof-Hutter}, {Shporer}, {Knutson}, {Liu}, {Thorngren}, {Lee}, {Chachan}, {Mawet}, {Millar-Blanchaer}, {Nilsson}, {Tinyanont}, {Vasisht}, \& {Wright}}]{Vissapragada2020}
{Vissapragada}, S., {Jontof-Hutter}, D., {Shporer}, A., {et~al.} 2020, \aj, 159, 108, \dodoi{10.3847/1538-3881/ab65c8}

\bibitem[{Vogt {et~al.}(1994)Vogt, Allen, Bigelow, Bresee, Brown, Cantrall, Conrad, Couture, Delaney, Epps, Hilyard, Hilyard, Horn, Jern, Kanto, Keane, Kibrick, Lewis, Osborne, Pardeilhan, Pfister, Ricketts, Robinson, Stover, Tucker, Ward, \& Wei}]{Vogt1994}
Vogt, S.~S., Allen, S.~L., Bigelow, B.~C., {et~al.} 1994, in Instrumentation in Astronomy VIII, ed. D.~L. Crawford \& E.~R. Craine, Vol. 2198, International Society for Optics and Photonics (SPIE), 362 -- 375, \dodoi{10.1117/12.176725}

\bibitem[{{Wachiraphan} {et~al.}(2025){Wachiraphan}, {Berta-Thompson}, {Diamond-Lowe}, {Winters}, {Murray}, {Zhang}, {Xue}, {Morley}, {Rosario-Franco}, \& {Duvvuri}}]{Wachiraphan2025}
{Wachiraphan}, P., {Berta-Thompson}, Z.~K., {Diamond-Lowe}, H., {et~al.} 2025, \aj, 169, 311, \dodoi{10.3847/1538-3881/adc990}

\bibitem[{{Wanderley} {et~al.}(2025){Wanderley}, {Cunha}, {Smith}, {Souto}, {Pascucci}, {Behmard}, {Allende Prieto}, {Beaton}, {Bizyaev}, {Daflon}, {Hasselquist}, {Howell}, {Majewski}, \& {Pinsonneault}}]{Wanderley2025}
{Wanderley}, F., {Cunha}, K., {Smith}, V.~V., {et~al.} 2025, arXiv e-prints, arXiv:2509.01930, \dodoi{10.48550/arXiv.2509.01930}

\bibitem[{{Weeks} {et~al.}(2025){Weeks}, {Van Eylen}, {Huber}, {Kawata}, {Stokholm}, {Aguirre B{\o}rsen-Koch}, {Pinilla}, {R{\o}rsted}, {Winther}, \& {Berger}}]{Weeks2025}
{Weeks}, A., {Van Eylen}, V., {Huber}, D., {et~al.} 2025, \mnras, 539, 405, \dodoi{10.1093/mnras/staf474}

\bibitem[{{Weiss} {et~al.}(2021){Weiss}, {Dai}, {Huber}, {Brewer}, {Collins}, {Ciardi}, {Matthews}, {Ziegler}, {Howell}, {Batalha}, {Crossfield}, {Dressing}, {Fulton}, {Howard}, {Isaacson}, {Kane}, {Petigura}, {Robertson}, {Roy}, {Rubenzahl}, {Twicken}, {Claytor}, {Stassun}, {MacDougall}, {Chontos}, {Giacalone}, {Dalba}, {Mocnik}, {Hill}, {Beard}, {Akana Murphy}, {Rosenthal}, {Behmard}, {Van Zandt}, {Lubin}, {Kosiarek}, {Lund}, {Christiansen}, {Matson}, {Beichman}, {Schlieder}, {Gonzales}, {Brice{\~n}o}, {Law}, {Mann}, {Collins}, {Evans}, {Fukui}, {Jensen}, {Murgas}, {Narita}, {Palle}, {Parviainen}, {Schwarz}, {Tan}, {Acton}, {Bryant}, {Chaushev}, {Gill}, {Eigm{\"u}ller}, {Jenkins}, {Ricker}, {Seager}, \& {Winn}}]{Weiss2021}
{Weiss}, L.~M., {Dai}, F., {Huber}, D., {et~al.} 2021, \aj, 161, 56, \dodoi{10.3847/1538-3881/abd409}

\bibitem[{{Wilson} {et~al.}(2003){Wilson}, {Eikenberry}, {Henderson}, {Hayward}, {Carson}, {Pirger}, {Barry}, {Brandl}, {Houck}, {Fitzgerald}, \& {Stolberg}}]{Wilson2003}
{Wilson}, J.~C., {Eikenberry}, S.~S., {Henderson}, C.~P., {et~al.} 2003, in Society of Photo-Optical Instrumentation Engineers (SPIE) Conference Series, Vol. 4841, Instrument Design and Performance for Optical/Infrared Ground-based Telescopes, ed. M.~{Iye} \& A.~F.~M. {Moorwood}, 451--458, \dodoi{10.1117/12.460336}

\bibitem[{{Wilson} {et~al.}(2018){Wilson}, {Teske}, {Majewski}, {Cunha}, {Smith}, {Souto}, {Bender}, {Mahadevan}, {Troup}, {Allende Prieto}, {Stassun}, {Skrutskie}, {Almeida}, {Garc{\'{\i}}a-Hern{\'a}ndez}, {Zamora}, \& {Brinkmann}}]{Wilson2018}
{Wilson}, R.~F., {Teske}, J., {Majewski}, S.~R., {et~al.} 2018, \aj, 155, 68, \dodoi{10.3847/1538-3881/aa9f27}

\bibitem[{{Wizinowich} {et~al.}(2000){Wizinowich}, {Acton}, {Shelton}, {Stomski}, {Gathright}, {Ho}, {Lupton}, {Tsubota}, {Lai}, {Max}, {Brase}, {An}, {Avicola}, {Olivier}, {Gavel}, {Macintosh}, {Ghez}, \& {Larkin}}]{wizinowich2000}
{Wizinowich}, P., {Acton}, D.~S., {Shelton}, C., {et~al.} 2000, \pasp, 112, 315, \dodoi{10.1086/316543}

\bibitem[{{Wordsworth} \& {Kreidberg}(2021)}]{Wordsworth2021}
{Wordsworth}, R., \& {Kreidberg}, L. 2021, arXiv e-prints, arXiv:2112.04663.
\newblock \doarXiv{2112.04663}

\bibitem[{{Xue} {et~al.}(2024){Xue}, {Bean}, {Zhang}, {Mahajan}, {Ih}, {Eastman}, {Lunine}, {Mansfield}, {Coy}, {Kempton}, {Koll}, \& {Kite}}]{Xue2024}
{Xue}, Q., {Bean}, J.~L., {Zhang}, M., {et~al.} 2024, \apjl, 973, L8, \dodoi{10.3847/2041-8213/ad72e9}

\bibitem[{{Yee} {et~al.}(2017){Yee}, {Petigura}, \& {von Braun}}]{Yee2017}
{Yee}, S.~W., {Petigura}, E.~A., \& {von Braun}, K. 2017, \apj, 836, 77, \dodoi{10.3847/1538-4357/836/1/77}

\bibitem[{{Zahnle} \& {Catling}(2017)}]{Zahnle2017}
{Zahnle}, K.~J., \& {Catling}, D.~C. 2017, \apj, 843, 122, \dodoi{10.3847/1538-4357/aa7846}

\bibitem[{{Zhang} {et~al.}(2024){Zhang}, {Hu}, {Inglis}, {Dai}, {Bean}, {Knutson}, {Lam}, {Goffo}, \& {Gandolfi}}]{Zhang2024}
{Zhang}, M., {Hu}, R., {Inglis}, J., {et~al.} 2024, \apjl, 961, L44, \dodoi{10.3847/2041-8213/ad1a07}

\bibitem[{{Zieba} {et~al.}(2023){Zieba}, {Kreidberg}, {Ducrot}, {Gillon}, {Morley}, {Schaefer}, {Tamburo}, {Koll}, {Lyu}, {Acu{\~n}a}, {Agol}, {Iyer}, {Hu}, {Lincowski}, {Meadows}, {Selsis}, {Bolmont}, {Mandell}, \& {Suissa}}]{Zieba2023}
{Zieba}, S., {Kreidberg}, L., {Ducrot}, E., {et~al.} 2023, \nat, 620, 746, \dodoi{10.1038/s41586-023-06232-z}

\bibitem[{{Zink} {et~al.}(2023){Zink}, {Hardegree-Ullman}, {Christiansen}, {Petigura}, {Boley}, {Bhure}, {Rice}, {Yee}, {Isaacson}, {Fernandes}, {Howard}, {Blunt}, {Lubin}, {Chontos}, {Pidhorodetska}, \& {MacDougall}}]{Zink2023}
{Zink}, J.~K., {Hardegree-Ullman}, K.~K., {Christiansen}, J.~L., {et~al.} 2023, \aj, 165, 262, \dodoi{10.3847/1538-3881/acd24c}

\end{thebibliography}
\bibliographystyle{aasjournal}

\end{document}